\documentclass[useAMS,usenatbib,usegraphicx]{mn2e}

\usepackage{pslatex,color,tabularx,float,multicol,caption}
\usepackage{amsmath,amsfonts,amssymb,amsxtra,eufrak,url}

\def\idm#1{{\mbox{\scriptsize #1}}}

\def\vec#1{{\pmb #1}}

\newcommand{\au}{\mbox{au}}

\newcommand{\mE}{\mbox{M}_{\oplus}}

\title[Migration of three planets and chains of MMRs]
{On the migration of three planets in a protoplanetary disc and the formation of chains of mean motion resonances}
\author[Cezary Migaszewski]{Cezary Migaszewski$^{1,2}$\thanks{E-mail:migaszewski@umk.pl}\\
$^{1}$Institute of Physics and CASA$^*$, Faculty of Mathematics and Physics, University of Szczecin, Wielkopolska 15, 70-451 Szczecin, Poland\\
$^{2}$Centre for Astronomy, Faculty of Physics, Astronomy and Informatics, Nicolaus Copernicus University, Grudziadzka 5, 87-100 Toru\'n, Poland}
\begin{document}
%
\date{Accepted 2016 February 16. Received 2016 January 25; in original form 2015 November 04}
\pagerange{\pageref{firstpage}--\pageref{lastpage}} \pubyear{2012}
\maketitle
\label{firstpage}

\captionsetup[figure]{labelfont=bf,font=small}

\begin{abstract}

We study the migration of three-planet systems in an irradiated 1+1D $\alpha$-disc with photoevaporation. We performed $2700$ simulations with various planets' masses and initial orbits. We found that most of the systems which ended up as compact configurations form chains of mean motion resonances (MMRs) of the first and higher orders. Most of the systems involved in chains of MMRs are periodic configurations. The period ratios of such system, though, are not necessarily close to exact commensurability. If a given system resides in a divergent migration zone in the disc, the period ratios increase and evolve along resonant divergent migration paths at ($P_2/P_1, P_3/P_2)$-diagram, where $P_1, P_2, P_3$ are the orbital periods of the first, second and third planet, respectively. The observed systems, though, do not lie on those paths. We show that an agreement between the synthetic and the observed systems distributions could be achieved if the orbital circularization was slower than it results from models of the planet-disc interactions. Therefore, we conclude that most of those systems unlikely formed as a result of divergent migration out of nominal chains of MMRs.

\end{abstract}


\section{Introduction}

It is widely accepted that planets form in protoplanetary discs and that gravitational interactions between the planets and the disc lead to planetary migration. Depending on the planets' masses and physical parameters of the disc, the migration can be inward or outward and the migration of a given two planets embedded in the disc can be convergent or divergent.

It is also well known that smooth convergent migration of two planets leads to resonant configurations \citep[e.g.,][]{Lee2002,Snellgrove2001,Papaloizou2005}. Moreover, if the migration is slow enough, a given system migrates in the phase space along families of periodic orbits \citep[e.g.,][]{Beauge2003,Beauge2006,Hadjidemetriou2006}. Although the migration of three planets has not been studied that extensively as for the two-planet system (in particular, the problem has not been studied in the context of periodic configurations), it is known that convergent migration leads to formation of chains of mean motion resonances \citep[e.g.,][]{Beauge2008,Cresswell2008,Papaloizou2010,Libert2011,Wang2012}.

If the slow, convergent migration was the dominant process in formation of planetary systems architectures, one should observe overpopulation of systems with period ratios close to the first- and higher-order mean motion resonances (MMRs). Nevertheless, such feature of the statistics of known multi-planet systems is not observed \citep{Fabrycky2014}.

Several mechanisms have been proposed to explain this discrepancy, i.e., stochastic forces acting on planets, resulting from turbulences in the disc \citep{Nelson2005,Rein2009} or the interaction with planetesimals \citep{Chatterjee2014}, energy dissipation due to planet-star tidal interactions \citep{Papaloizou2010,Papaloizou2011,Batygin2013,Delisle2014a,Delisle2014b} or planet-disc wake interaction \citep{Podlewska-Gaca2012,Baruteau2013}.

In our recent work \citep[][Paper~I from hereafter]{Migaszewski2015}, we showed that in a standard 1+1D irradiated $\alpha$-disc model with a realistic opacity law there exist zones of convergent as well as divergent migration. During the disc evolution, positions and sizes of the zones change, which makes the evolution of a given system complex. Even if the system got trapped in a resonance (in a sense of the period ratio), the period ratio increases when the system resides in the divergent migration zone. 

In this paper we extend the analysis in Paper~I, devoted to two-planet systems, to a three-planet case. In Paper~I we showed that almost all the synthetic systems with $P_2/P_1 \lesssim 2.12$ are resonant in terms of the librating critical angles, even if their period ratios are distant from the nominal values of MMRs (only first order resonances were present in the sample of the synthetic systems). Another common feature of those systems is that their evolution is periodic after the disc disperses.

\begin{figure*}
\begin{center}
\includegraphics[width=0.7\textwidth]{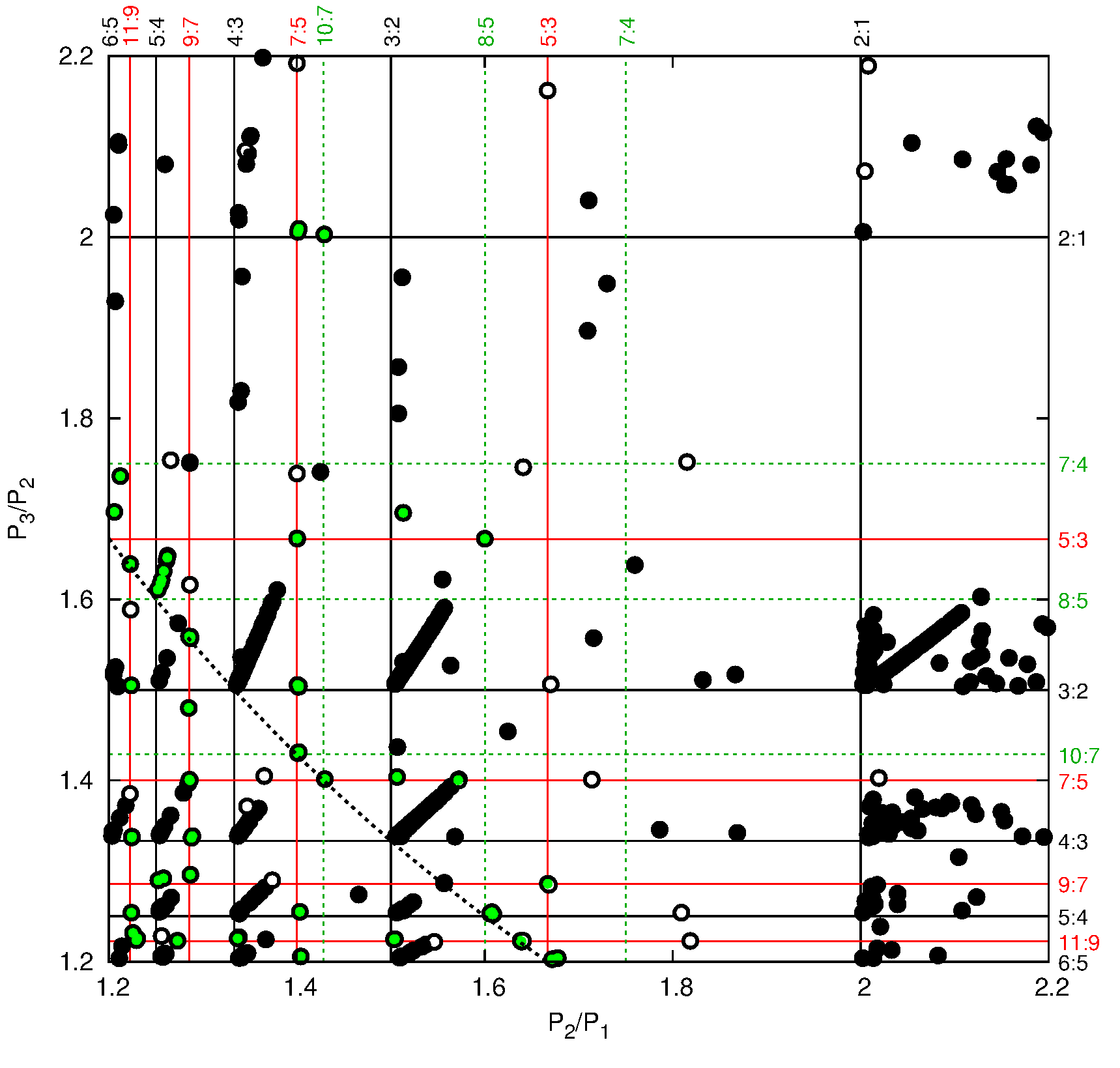}
\end{center}
\caption{Final systems resulting from the migration simulations presented at the $(P_2/P_1, P_3/P_2)-$plane. Only systems with both $P_2/P_1$ and $P_3/P_2$ smaller than $2.2$ are shown. Black filled circles are for systems with both pairs of planets involved in the first-order resonances (i.e., at least one resonant angle for each pair of planets librates). Green circles show configurations with both pairs resonant, but at least one pair is involved in higher-order MMR. Empty circles show systems which do not form chains of MMRs, although one of the pairs may be involved in MMR. Vertical and horizontal lines show positions of MMRs (labelled accordingly) for the first and the second pair of planets, respectively. Black solid lines denote the first-order MMRs, red solid lines are for the second-order MMRs, green dashed lines mark positions of the third-order MMRs. Black dashed curve denote 2:1~MMR between the innermost and the outermost planet.}
\label{fig:stat1}
\end{figure*}

We show that the conclusions are similar for systems with three super-Earth mass planets. We used the same model of the disc and performed $2700$ simulations for various planets' masses and initial orbits. Most of the systems with period ratios below $\sim 2.12$ end up as resonant configurations (chains of MMRs), also when the period ratios are significantly different from the exact commensurability. Unlike in two-planet systems, higher order resonances are also present in the sample. Similarly to the two-planet case, most of those systems are periodic configurations.

We found that if a given system is resonant in terms of librating resonant angles (and the orbit circularization due to planet-disc interaction is much faster than the migration, as it is in our disc model), the divergent migration of the system occurs along certain paths at the period ratio -- period ratio diagram (which we call the resonant divergent migration paths). It is consistent with results of \cite{Papaloizou2015} and~\cite{Batygin2013} who studied the divergent migration of three-planet systems under the tidal star-planet interactions. 

Majority of the synthetic systems in our simulations, with period ratios below or around $2$, lie along those paths, which is not consistent with the observed three-planet systems. We show that slower circularization (with respect to generally accepted rates) could cause a system migrating divergently to leave the path, which could make the statistics of synthetic systems more alike the observational sample.

The paper is organized as follows. In Section~2 we shortly overview the model and present the parameters input. Section~3 is devoted to the presentation of the simulation results. In Section~4 we compare the results with the observational sample of three-planet systems. In the last section we gather the conclusions.

\section{The model and the parameters input}

We use the disc model described in Paper~I. It is a standard 1+1D $\alpha$-disc model \citep[e.g.,][]{Garaud2007} with the stellar irradiation \citep{Ruden1991} and the photoevaporation \citep{Matsuyama2003}. The value of $\alpha=0.004$ was chosen such that the disc life time of $\sim 3.5~$Myr is appropriate to make a few-Earth mass planets, which start from $\sim 1\,\au$, migrate inwards down to a few-day orbits. The opacity law is taken from \citep{Semenov2003}.

\begin{figure*}
\begin{center}
\vbox{
\hbox{
\includegraphics[width=0.48\textwidth]{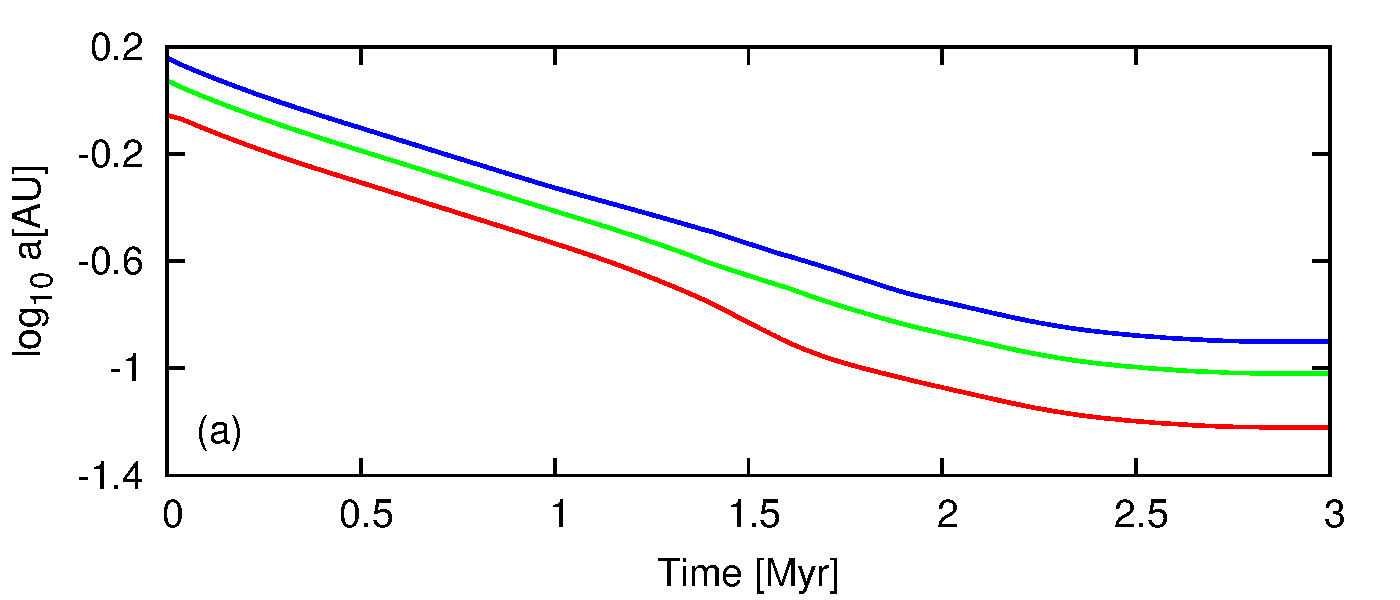}
\includegraphics[width=0.48\textwidth]{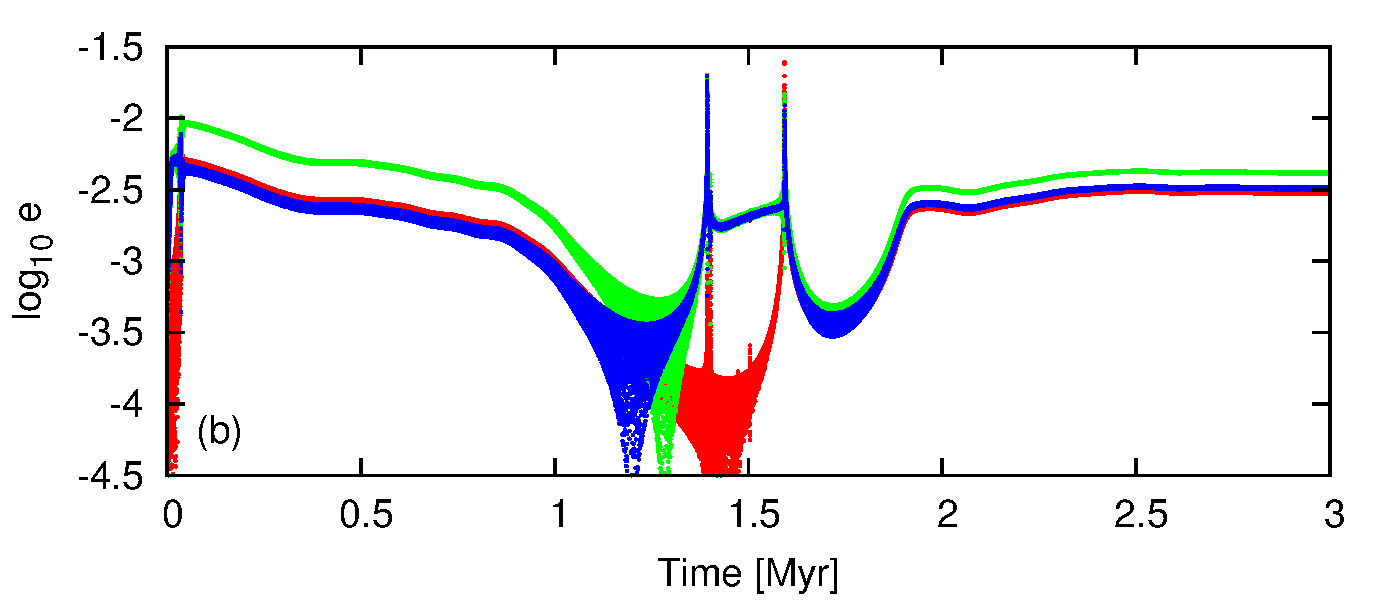}
}
\hbox{
\includegraphics[width=0.48\textwidth]{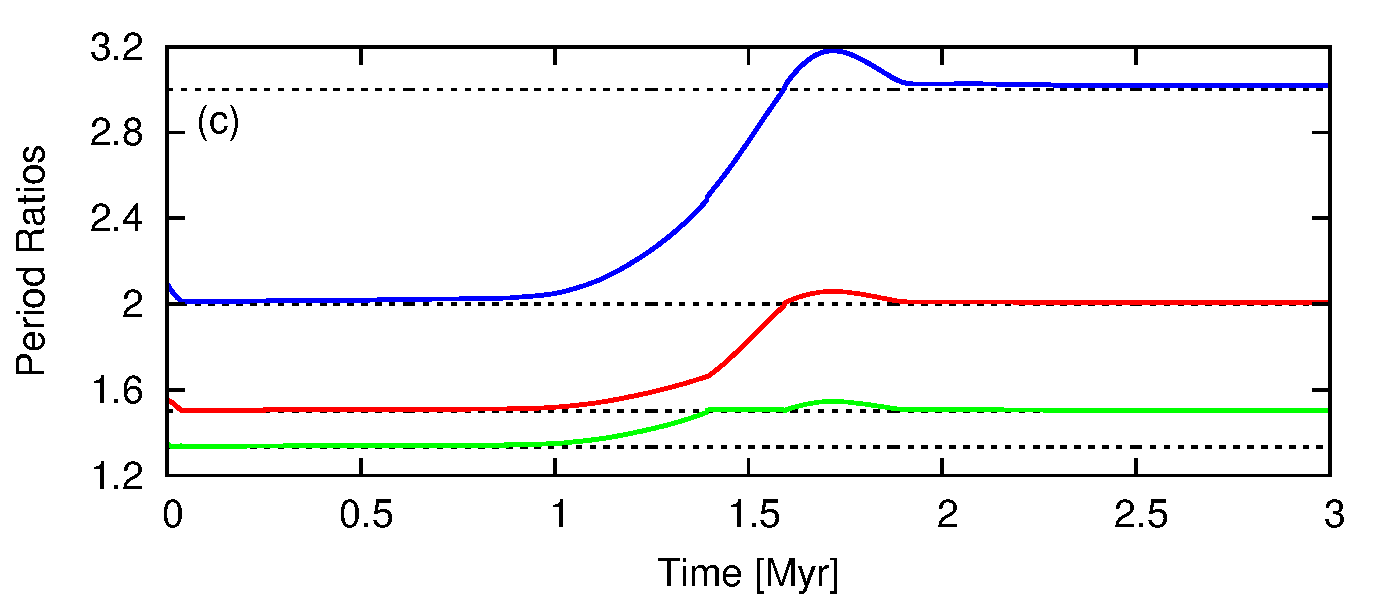}
\includegraphics[width=0.48\textwidth]{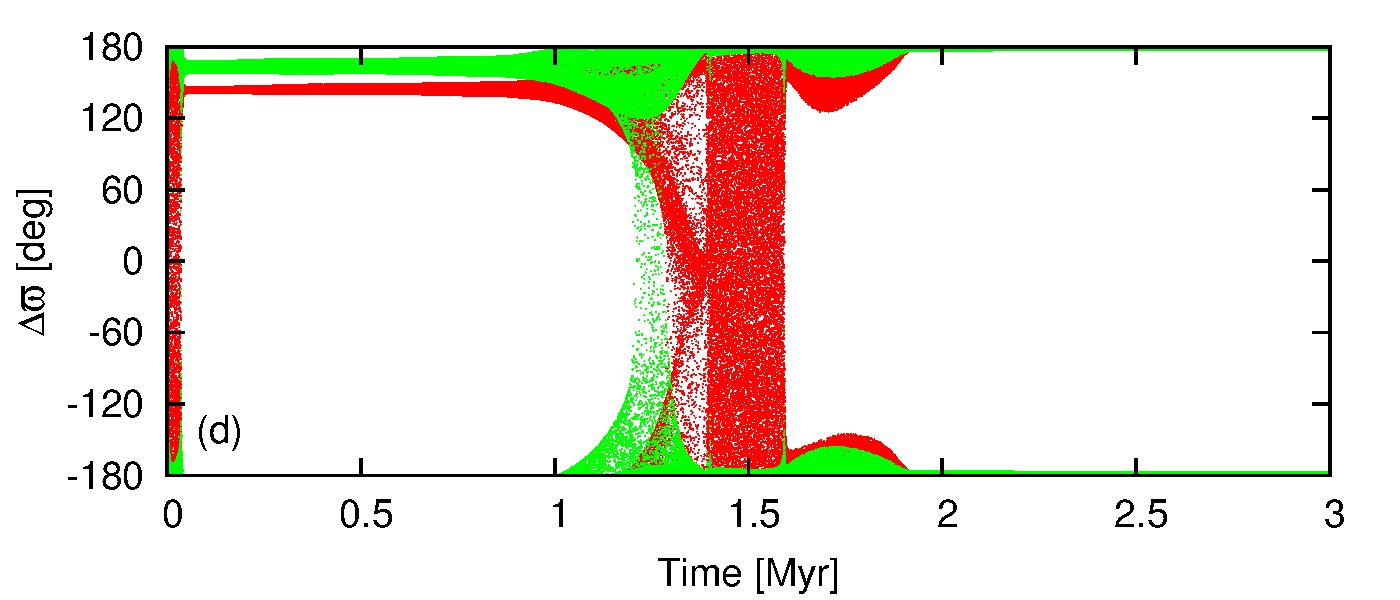}
}
\hbox{
\includegraphics[width=0.48\textwidth]{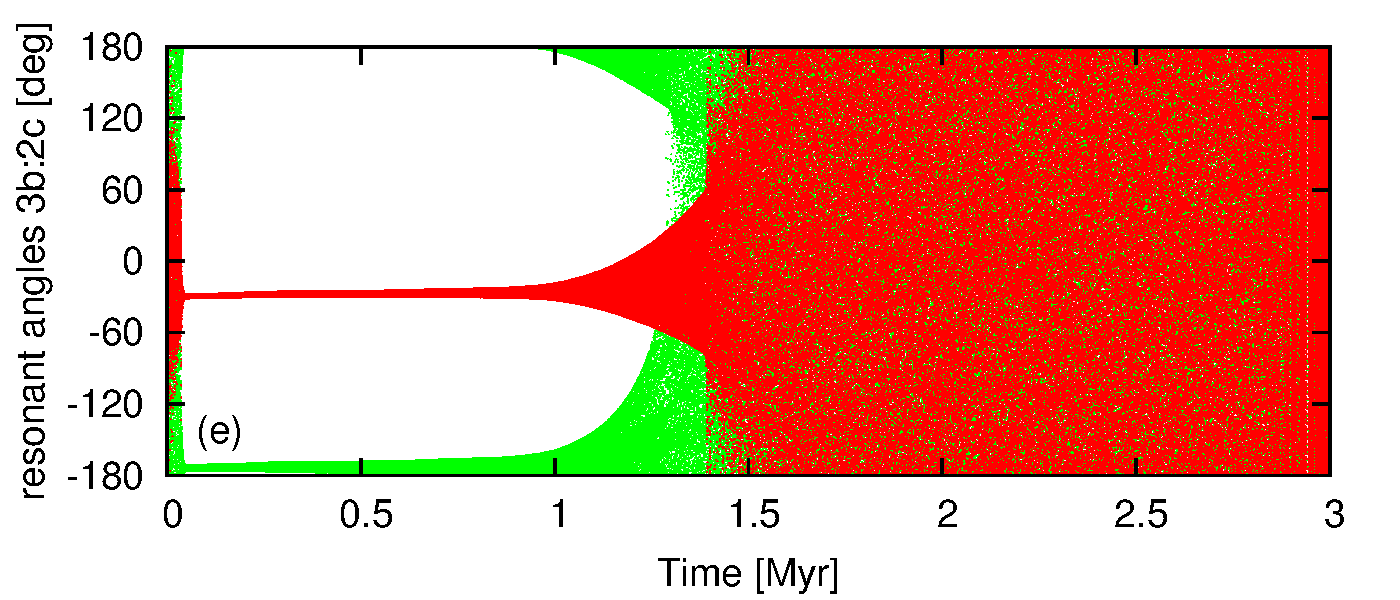}
\includegraphics[width=0.48\textwidth]{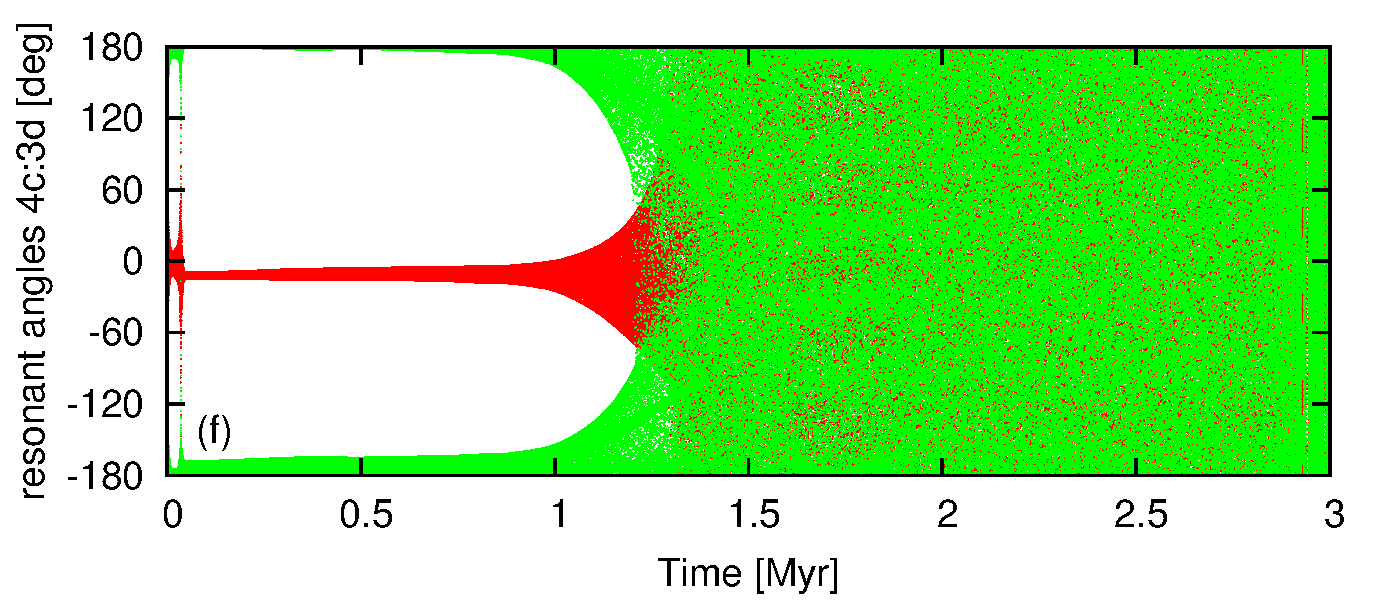}
}
\hbox{
\includegraphics[width=0.48\textwidth]{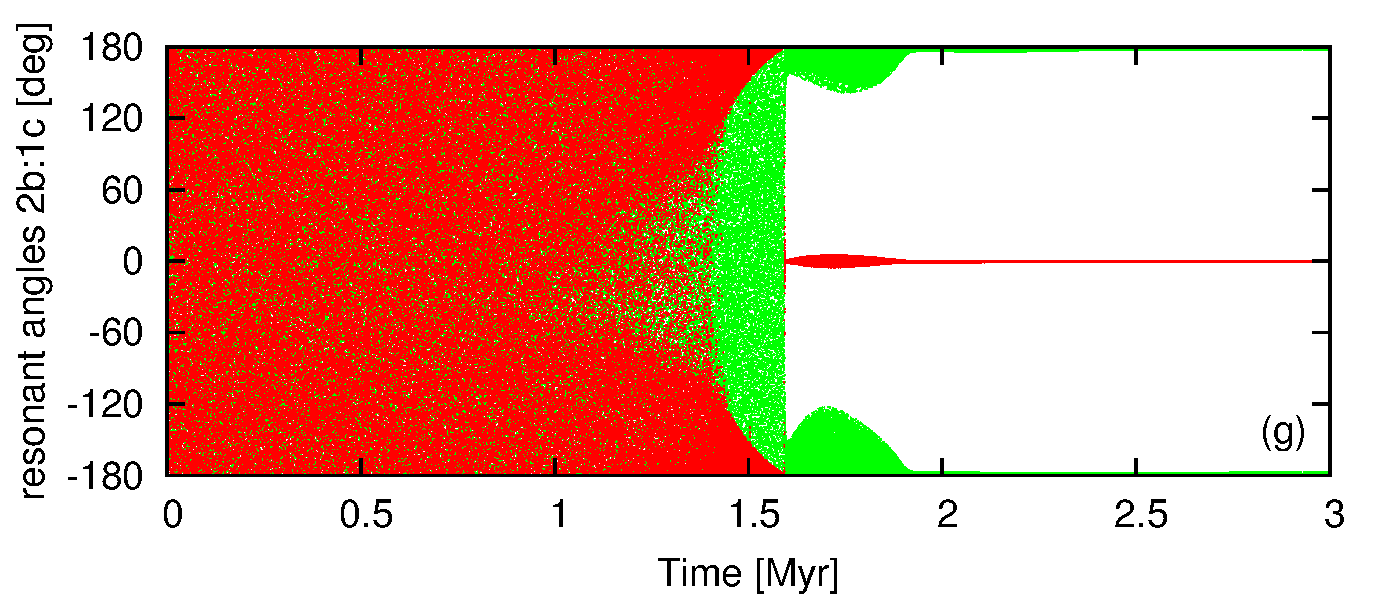}
\includegraphics[width=0.48\textwidth]{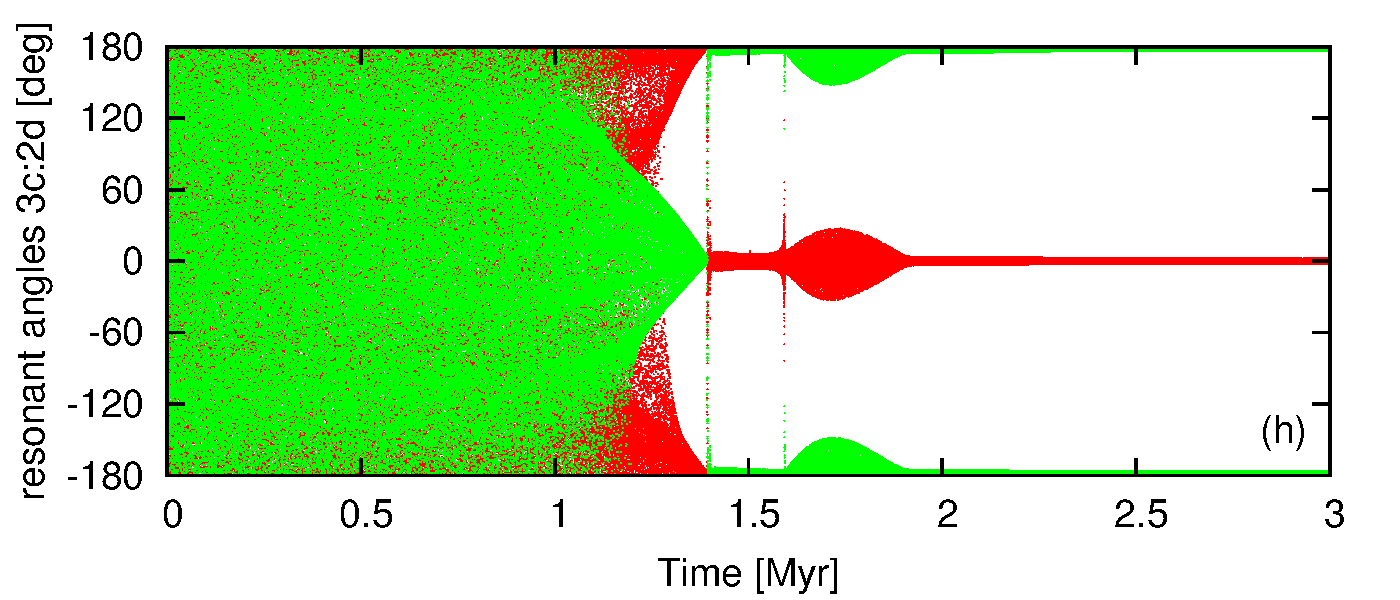}
}
}
\end{center}
\caption{The evolution of an example system presented at different planes. Panels~(a) and~(b) show the evolution of semi-major axes and eccentricities, respectively. Colours red (grey in printed version), green (light grey) and blue (dark grey) are for the first, second and third planet, respectively. Panel~(c) presents the evolution of the period ratios, $P_2/P_1$ (red/grey), $P_3/P_2$ (green/light grey) and $P_3/P_1$ (blue/dark grey). Panel~(d) shows the evolution of $\Delta\varpi_{1,2} \equiv \varpi_1-\varpi_2$ (red/grey) and $\Delta\varpi_{2,3} \equiv \varpi_2-\varpi_3$ (green/light grey). Remaining panels present the evolution of the critical angles of MMRs (see labels at each panel). For each panel red/grey and green/light grey colours are for the angles with the pericenter longitude of the inner and the outer planet of a given pair in its definition, respectively. Planets' masses are $m_1=4.940\,\mE$, $m_2=4.447\,\mE$ and $m_3=4.876\,\mE$.}
\label{fig:ex1}
\end{figure*}

The planet-disc interaction is being computed by using analytic prescriptions for the Lindblad and the corotation torques \citep{Paardekooper2011}. The eccentricity damping is governed by the eccentricity waves \citep{Tanaka2004}. The model also accounts for the dependence of the corotation torque on the eccentricity \citep{Fendyke2014} as well as formulae for the transition between type~I and type~II migration \citep{Dittkrist2014}, which is important even for low mass planets when the disc is not massive in  late stages of its evolution. We also account for the axially symmetric potential of the disc which contributes to the rates of the periapsis rotation. The details of the disc model as well as the planet-disc interactions treatment can be found in Paper~I.

We performed $2700$ simulations for the planets' masses chosen randomly from a range of $[3, 5]\,\mE$. The initial semi-major axis of the innermost planet $a_1$ was being chosen from a range of $\log_{10}a_1[\au] \in [-0.1, 0.3]$, while the period ratios for both pairs of planets, $P_2/P_1$ and $P_3/P_2$ -- from ranges of $[1.2, 1.8]$ and $[1.2, 2.2]$ for two different sets of simulations. The evolution of the planets starts at $t=0.5$~Myr of the disc evolution and are followed up to $3.5$~Myr. In the next section we present the statistical properties of the sample of the final configurations.

\section{Statistics of the synthetic systems}

Figure~\ref{fig:stat1} presents final systems which ended up with both $P_2/P_1$ and $P_3/P_2$ below $2.2$ at the $(P_2/P_1, P_3/P_2)-$plane (there are $\sim 1000$ such  configurations). Because of the existence of the planetary trap in the disc at $a \sim 1\,\au$ (which is an orbit at which the total torque acting on a planet vanishes; e.g., Paper~I), for some of the systems final $P_2/P_1$ or $P_3/P_2$ are larger than $2.2$. Vertical and horizontal lines mark positions of MMR (solid black, solid red and dashed green for the first-, second- and third-order MMRs, respectively). Points defined by a vertical and a horizontal lines crossing one another denote nominal positions of chains of MMRs at the diagram. For instance a point of coordinates $(3/2, 4/3)$ means a nominal chain of 3:2 and 4:3~MMRs for the inner and the outer pair of planets, respectively. This particular chain of resonances leads to 2:1~MMR between the innermost and the outermost planets. A black dashed curve shows the position of this resonance at the diagram.

A configuration is called a chain of MMRs if both the pairs are resonant and a given pair of planets is called resonant if at least one of its resonant angles librates. Similarly to the two-planet systems resulting from the smooth migration (Paper~I), the three-planet systems can be resonant in wide ranges of the period ratios. It means that most of the systems with both $P_2/P_1$ and $P_3/P_2$ smaller than $2.2$ are chains of MMRs. Black filled circles mark configurations which are chains of the first-order MMRs, while green symbols indicate that at least one of the resonant pairs is involved in higher-order MMR. Empty circles denote positions of systems, which are not chains of MMRs, although one of the pairs may be resonant.

A characteristic feature of this diagram is that majority of systems are placed along skew lines (they are actually curves) originating from positions of the nominal chains of MMR (mainly of the first order). As we will show later, those systems resided close to the nominal chains of MMRs but underwent the divergent migration later. We call those characteristic curves the resonant divergent migration paths, because a system which migrates divergently as a resonant configuration moves along one of those paths at the period ratio -- period ratio diagram.

\subsection{Chains of mean motion resonances}

In principle, the migration of a given system may be very complex. The system can move between the convergent the divergent migration zones a few times. The final state of the system depends on the initial orbits and planets' masses. In this and subsequent sections we discuss the evolution of two interesting examples, which are representative for systems forming chains of MMRs. The first system, whose evolution is presented below, ends up the migration as a chain of first-order resonances.

\begin{figure}
\begin{center}
\vbox{
\hbox{
\includegraphics[width=0.48\textwidth]{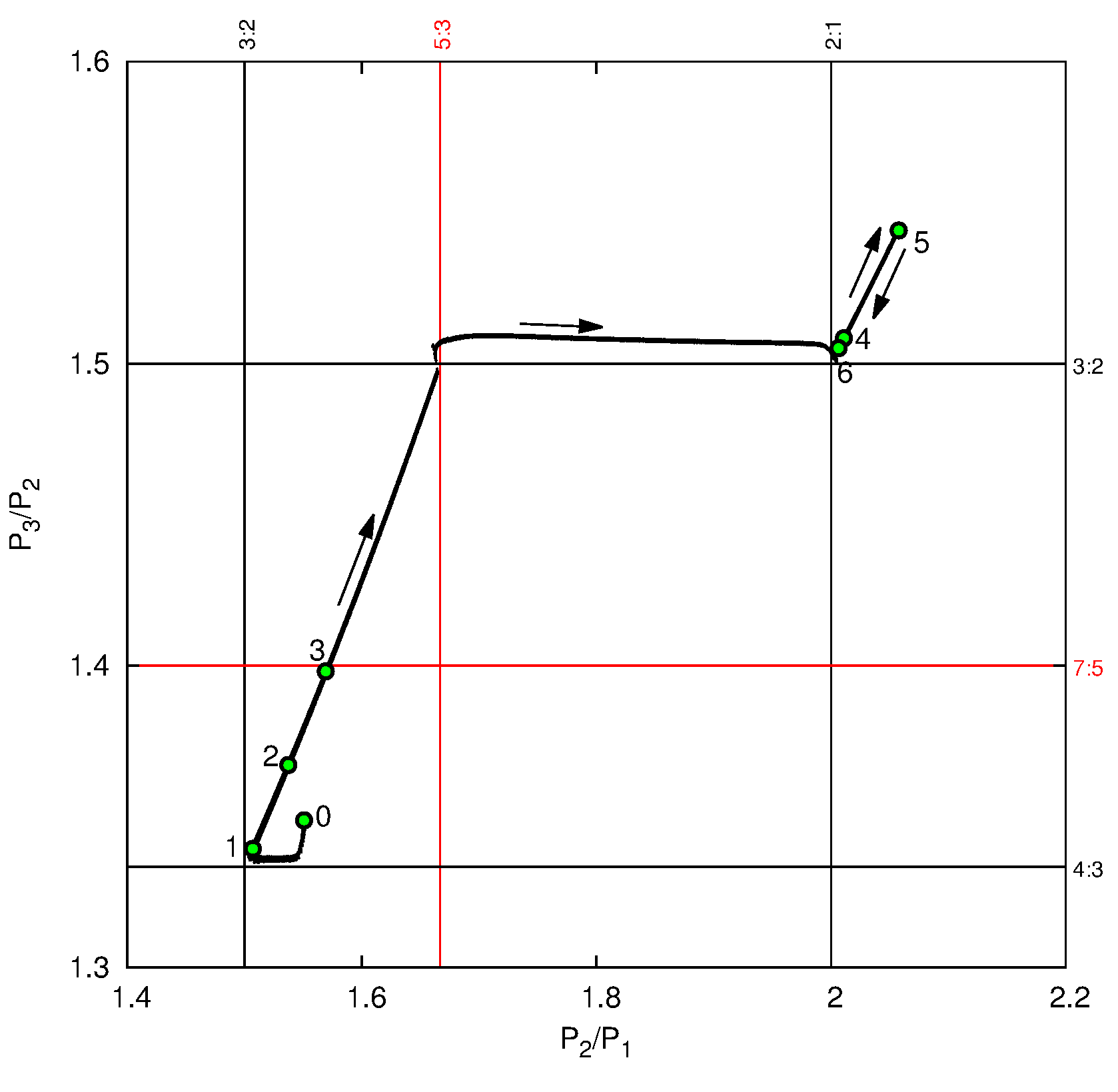}
}
}
\end{center}
\caption{The evolution (a black curve) of the system illustrated in Fig.~\ref{fig:ex1}, here presented at the period ratio -- period ratio diagram. Green dots (grey in the printed version) show the positions of the system in chosen moments of the evolution. Subsequent labels from $0$ to $6$ mean $t=0, 0.5, 1.1, 1.2, 1.7, 1.9$ and~$3.0~$Myr. Horizontal and vertical lines indicate the positions of MMRs (labelled accordingly).}
\label{fig:ex1a}
\end{figure}

Figure~\ref{fig:ex1} shows the evolution of this system at several planes (see the labels at each panel). The system starts close to a chain of 3:2 and 4:3 MMRs (see Fig.~\ref{fig:ex1}c, the evolution of the system can be also followed at the period ratio -- period ratio diagram, Fig.~\ref{fig:ex1a}). It reaches the chain and remains in it for some time. Both critical angles of 3:2 and 4:3~MMRs librate. The first pair of planets is involved in 3:2~MMR  (see Fig.~\ref{fig:ex1}e), thus the critical angles are $\phi_{1,1} = 2\,\lambda_1 - 3\,\lambda_2 + \varpi_1$ and $\phi_{1,2} = 2\,\lambda_1 - 3\,\lambda_2 + \varpi_2$. The first angle (red colour) librates around a value close to $0$, while the second one (green colour) -- close to $180$~degrees. Similar situation occurs for the critical angles of the second pair of planets (see Fig.~\ref{fig:ex1}f). The angles are $\phi_{2,1} = 3\,\lambda_2 - 4\,\lambda_3 + \varpi_2$ (red colour) and $\phi_{2,2} = 3\,\lambda_2 - 4\,\lambda_3 + \varpi_3$ (green colour). As all the critical angles librate, the differences between apsidal lines also librate (see Fig.~\ref{fig:ex1}d, the evolution of $\Delta\varpi_{1,2} \equiv \varpi_1 - \varpi_2$ and $\Delta\varpi_{2,3} \equiv \varpi_2 - \varpi_3$ is plotted in red and green, respectively). Moreover, as two pairs of planets $1$ and $2$ as well as $2$ and $3$ are involved in 3:2 and 4:3~MMRs, planets $1$ and $3$ are involved in 2:1~MMR. The critical angles of this resonance also librate (it is not shown here).

After slightly more than $1~$Myr both period ratios $P_2/P_1$ and $P_3/P_2$ start to deviate from the nominal values of the resonances. The divergent migration of both pairs of planets is linked (what will be explained later) until $P_3/P_2$ reaches a nominal value of 3:2~MMR. After that the evolution of the period ratios is not linked any more, i.e., they evolve independently. The period ratio of the inner pair keeps increasing, while the outer pair reaches 3:2~MMR and $P_3/P_2$ stays close to the nominal value of this MMR until $P_2/P_1$ reaches the nominal value of 2:1~MMR. Then the migration of two pairs of planets is linked again. Both $P_2/P_1$ and $P_3/P_2$ increase together for a short time, after which they decrease again reaching $2:1$ and $3:2$~MMR, respectively. This chain of resonances is the final stage of the system. Both resonant angles of each pair librate. For the first pair of planets the critical angles are $\phi_{1,1} = \lambda_1 - 2\,\lambda_2 + \varpi_1$ (red colour in Fig.~\ref{fig:ex1}g) and $\phi_{1,2} = \lambda_1 - 2\,\lambda_2 + \varpi_2$ (green colour in Fig.~\ref{fig:ex1}g). For the second pair of planets the angles read $\phi_{2,1} = 2\,\lambda_2 - 3\,\lambda_3 + \varpi_2$ (red colour in Fig.~\ref{fig:ex1}h) and $\phi_{2,2} = 2\,\lambda_2 - 3\,\lambda_3 + \varpi_3$ (green colour in Fig.~\ref{fig:ex1}h).

\begin{figure*}
\begin{center}
\vbox{
\hbox{
\includegraphics[width=0.33\textwidth]{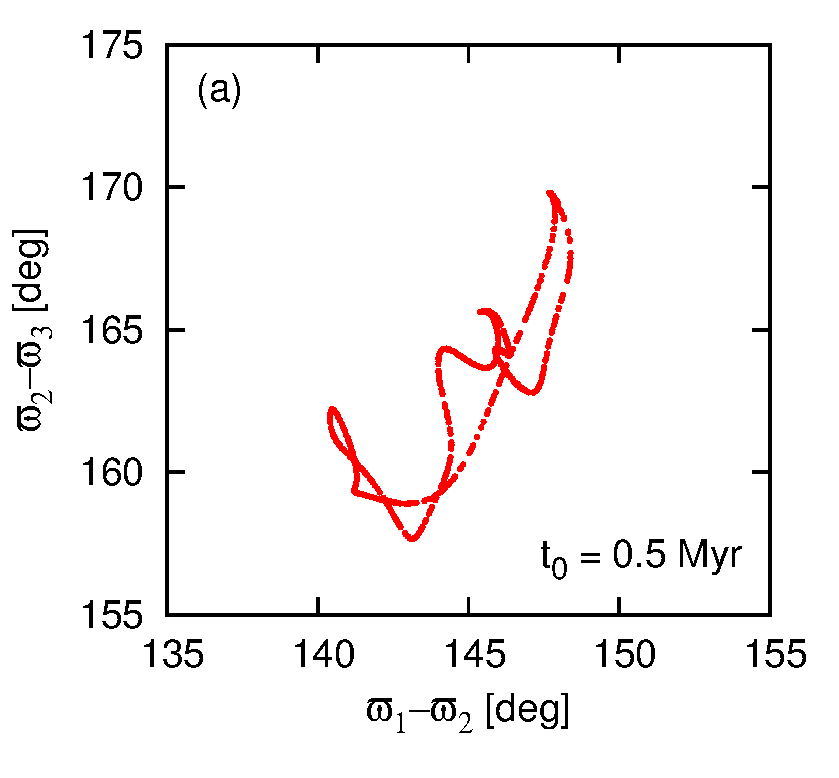}
\includegraphics[width=0.33\textwidth]{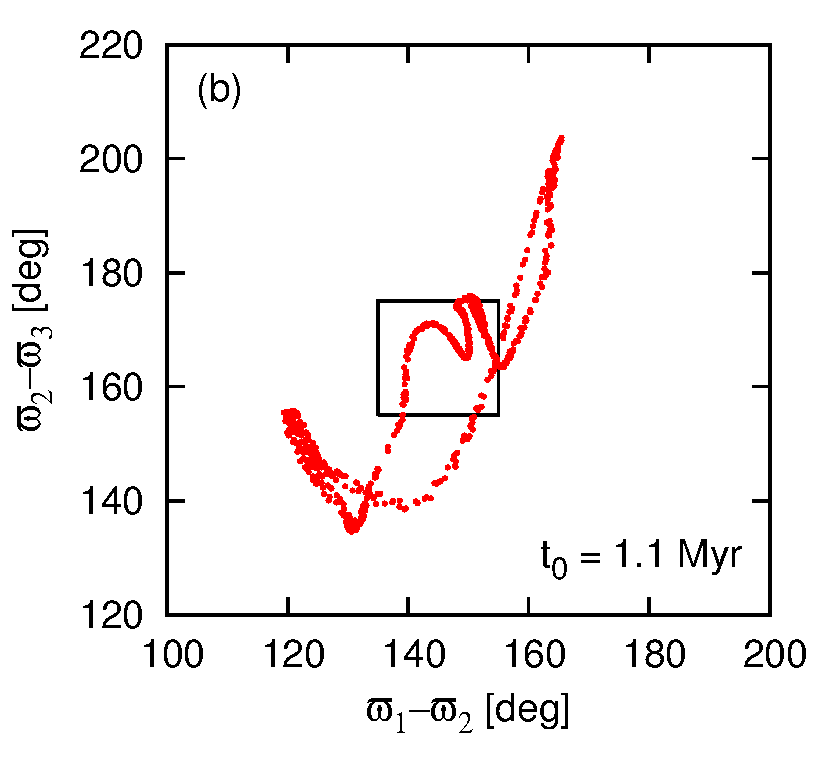}
\includegraphics[width=0.33\textwidth]{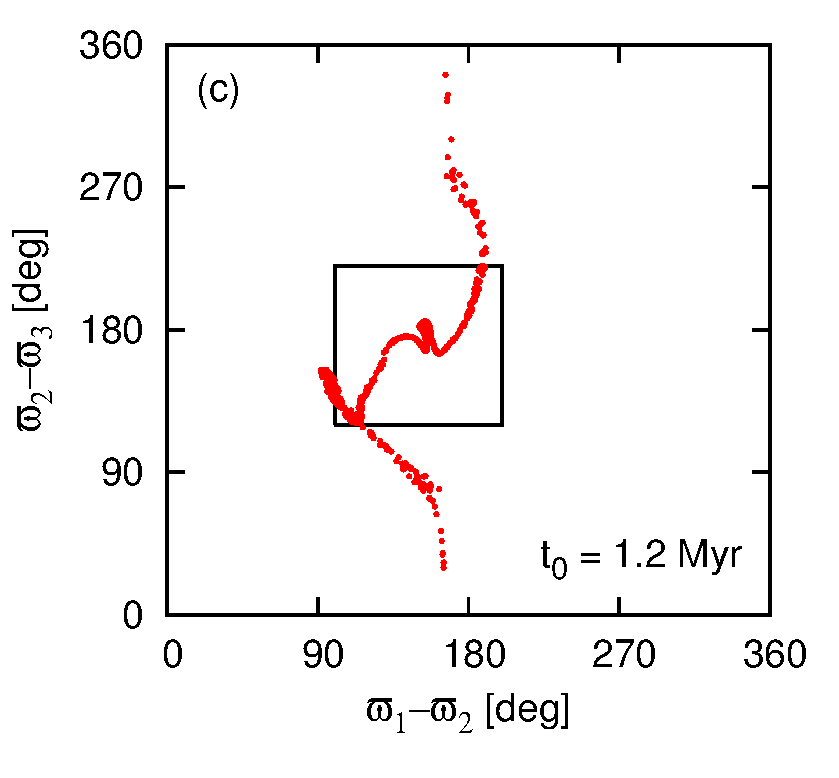}
}
\hbox{
\includegraphics[width=0.33\textwidth]{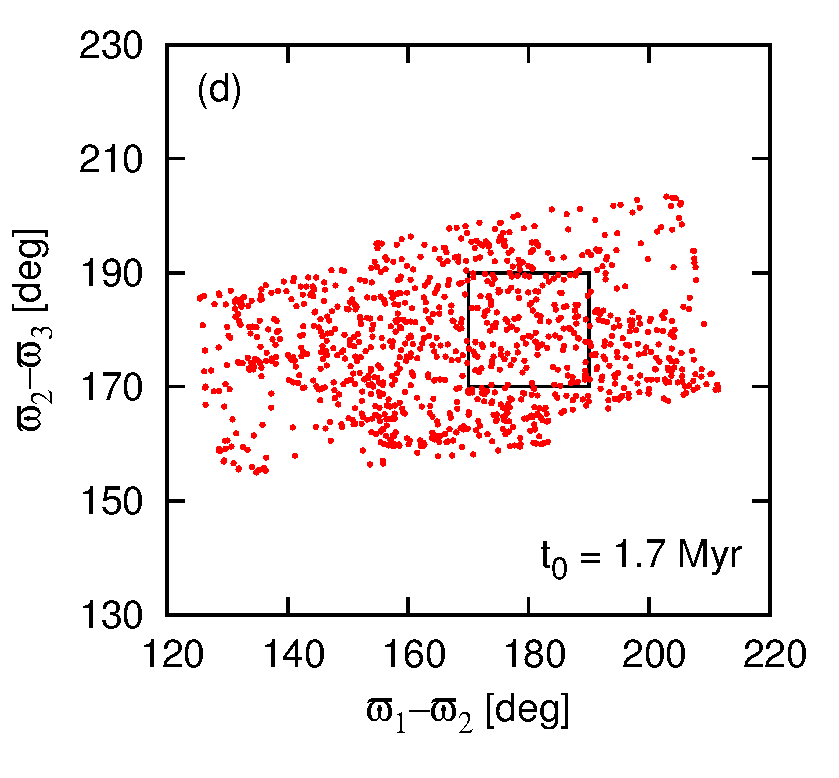}
\includegraphics[width=0.33\textwidth]{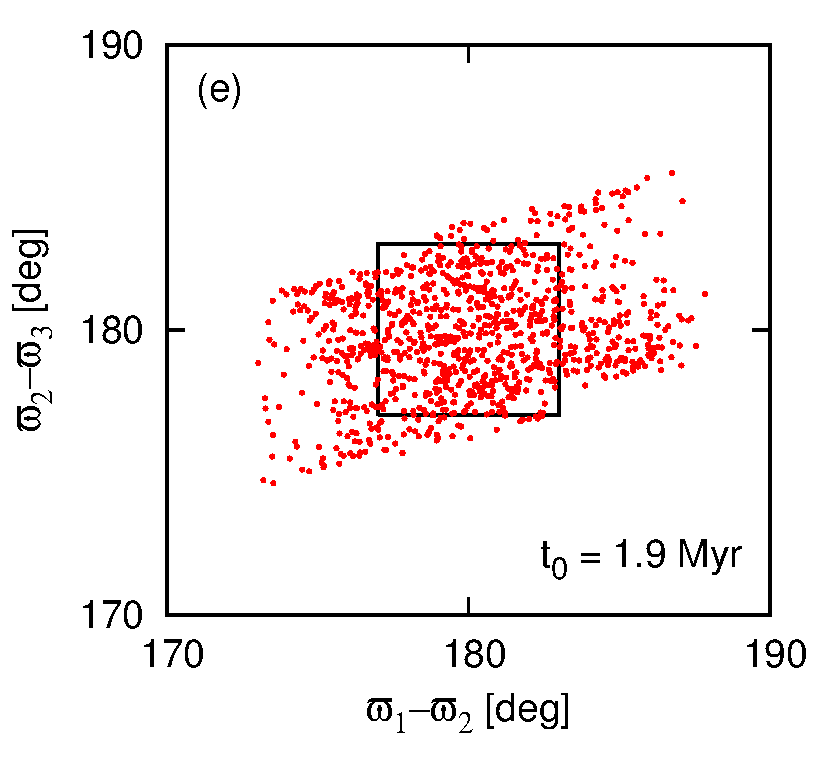}
\includegraphics[width=0.33\textwidth]{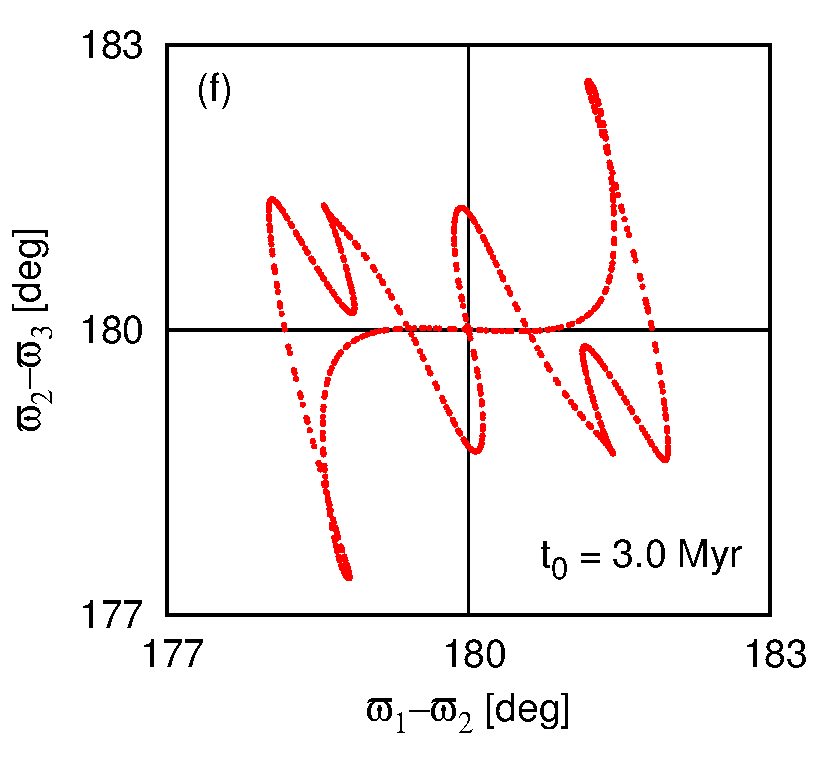}
}
}
\end{center}
\caption{The evolution of the system illustrated in Fig.~\ref{fig:ex1} shown at $(\Delta\varpi_{1,2}, \Delta\varpi_{2,3})-$planes. Each plot presents the evolution for a time interval $t \in [t_0 - \delta\,t, t_0 + \delta\,t]$, where $t_0$ is given at each panel and $\delta\,t = 10$~kyr. The positions of the system at the $(P_2/P_1, P_3/P_2)$-plane in those moments of time are marked with green dots in Fig.~\ref{fig:ex1a}. Black squares show ranges of the previous or the next plot in the sequence, showing the changes of the libration amplitudes of $\Delta\varpi_{1,2}$ and $\Delta\varpi_{2,3}$, see the text for details.}
\label{fig:periodic1}
\end{figure*}

\subsection{Periodic orbits}

The evolution of the system discussed in the previous section can be also studied at the $(\Delta\varpi_{1,2}, \Delta\varpi_{2,3})-$plane. A system which formed a chain of MMRs via convergent migration becomes a periodic configuration. Even if at later time it undergoes the divergent migration, the evolution remains periodic, until the system reaches other MMRs. Figure~\ref{fig:periodic1} shows the positions of the system at the $(\Delta\varpi_{1,2}, \Delta\varpi_{2,3})-$plane for $t \in [t_0 - \delta\,t, t_0 + \delta\,t]$ for six different epochs $t_0$ and $\delta\,t = 10$~kyr. Figure~\ref{fig:periodic1}a shows the evolution for $t_0 = 0.5~$Myr, when the system resides in the chain of 3:2 and 4:3~MMR (see Fig.~\ref{fig:ex1}c). The position of the system at the period ratio -- period ratio diagram is marked with a green dot labelled with $1$ in Fig.~\ref{fig:ex1a}. All the critical angles librate with small amplitudes, therefore also $\Delta\varpi_{1,2}$ and $\Delta\varpi_{2,3}$ librate with small amplitudes of $\sim 10~$degrees. We can call the evolution periodic, nevertheless, because of the dissipative forces acting on the planets, it is not periodic in the common sense. Periodic orbits form a family of trajectories in the phase space. In a problem of three planets the family is parametrized with one of the period ratio. When a system migrates divergently, the period ratios increase, thus at each epoch the system represents a different periodic configuration of the family. We can say that the migration occurs along the family of periodic orbits. The evolution of a given orbital parameter $\theta_j(t)$ is not strictly periodic, i.e., after the period $T$, $\theta_j(t + T) = \theta_j(t) + \delta\theta_j$, where the last term stems from the dissipative evolution of the system and its value is small if the migration is slow.

After some time (at $t=1.1~$Myr, see also Fig.~\ref{fig:ex1a} for the positions of the system at the period ratio -- period ratio diagram) the system starts to deviate from the nominal values of the MMRs. The period ratios increase and so do the amplitudes of the librations of the resonant angles and $\Delta\varpi_{i,i+1}$, $i=1,2$, (Fig.~\ref{fig:periodic1}b). A black square shows the $x-$ and $y-$ranges of the previous panel (Fig.~\ref{fig:periodic1}a) to illustrate the increase of the amplitudes of $\Delta\varpi_{i,i+1}$ variation. The dissipative evolution is faster, thus the $\delta\theta_j$ term is larger, nevertheless, the evolution remains periodic in the sense described above. Further divergent evolution leads to rotation of $\Delta\varpi_{2,3}$ (Fig.~\ref{fig:periodic1}c). Both period ratios are below the values of first-order MMRs located above the initial MMRs (2:1~MMR for the pair migrating away from 3:2~MMR and 3:2~MMR for the pair migrating away from 4:3~MMR) but the second angle of the 3:2~MMR already started to rotate. It happens because the system reached 7:5~MMR for the second pair (see a green point labelled with $3$ in Fig.~\ref{fig:ex1a}). Nevertheless, the evolution is still periodic. 

The next panel (Fig.~\ref{fig:periodic1}d) shows the evolution of $\Delta\varpi_{i,i+1}$ for $t_0 = 1.7~$Myr. This moment of the evolution corresponds to the period ratios above 2:1 (for the inner pair) and 3:2~MMR (for the outer pair) and further divergent migration. Even though the system is resonant with all resonant angles librating, the evolution is not periodic. After analysing the evolution of other systems in the sample, we conclude that in order to establish the periodic configuration, a given system needs to migrate convergently and the period ratios have to approach to the nominal values of MMRs for both pairs of planets. In the situation discussed here one of the above conditions is not fulfilled, as the migration was divergent before the system reached the 2:1, 3:2 chain of MMRs and remained divergent after the system passed through this chain. After some time the migration becomes convergent (see Fig.~\ref{fig:periodic1}e, for $t_0 = 1.9~$Myr) but the evolution is still not periodic, because the period ratios are still far from the nominal values. Finally, after the period ratios reach the nominal values of MMRs, the evolution becomes periodic again. The last panel of Fig.~\ref{fig:periodic1} shows the evolution of the system at the end of the simulation. Because the disc is dispersed, there is no dissipation and the evolution is strictly periodic ($\delta\theta_j = 0$).

\begin{figure*}
\begin{center}
\vbox{
\hbox{
\includegraphics[width=0.48\textwidth]{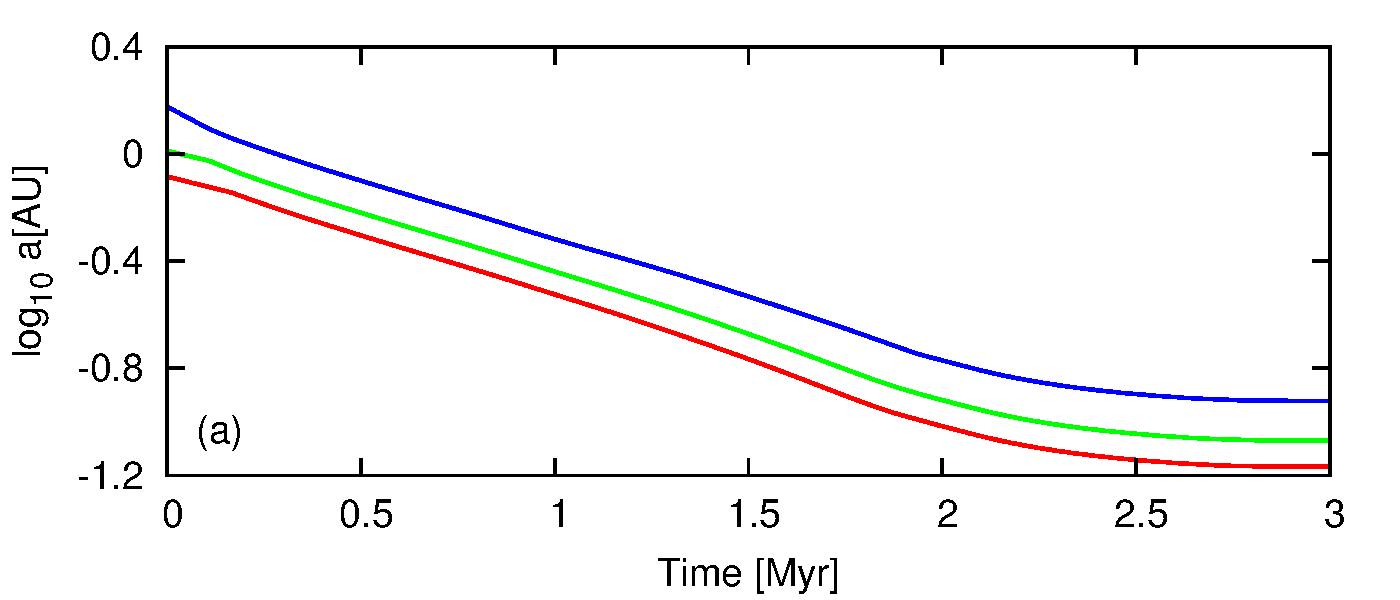}
\includegraphics[width=0.48\textwidth]{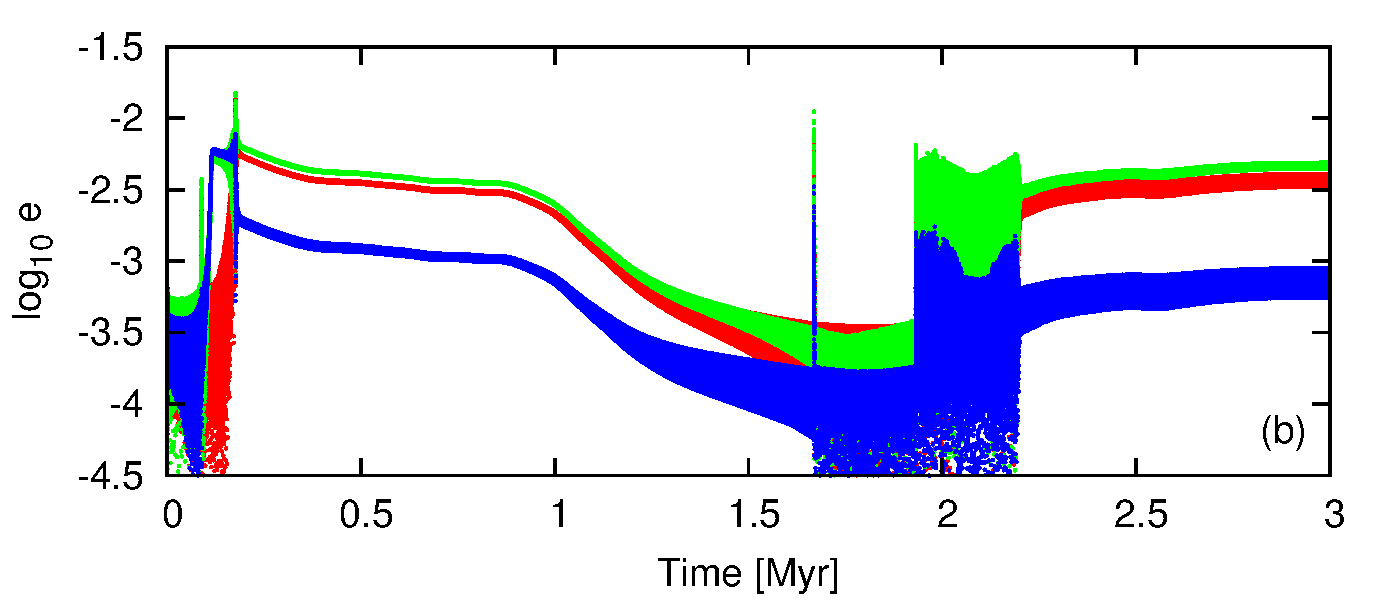}
}
\hbox{
\includegraphics[width=0.48\textwidth]{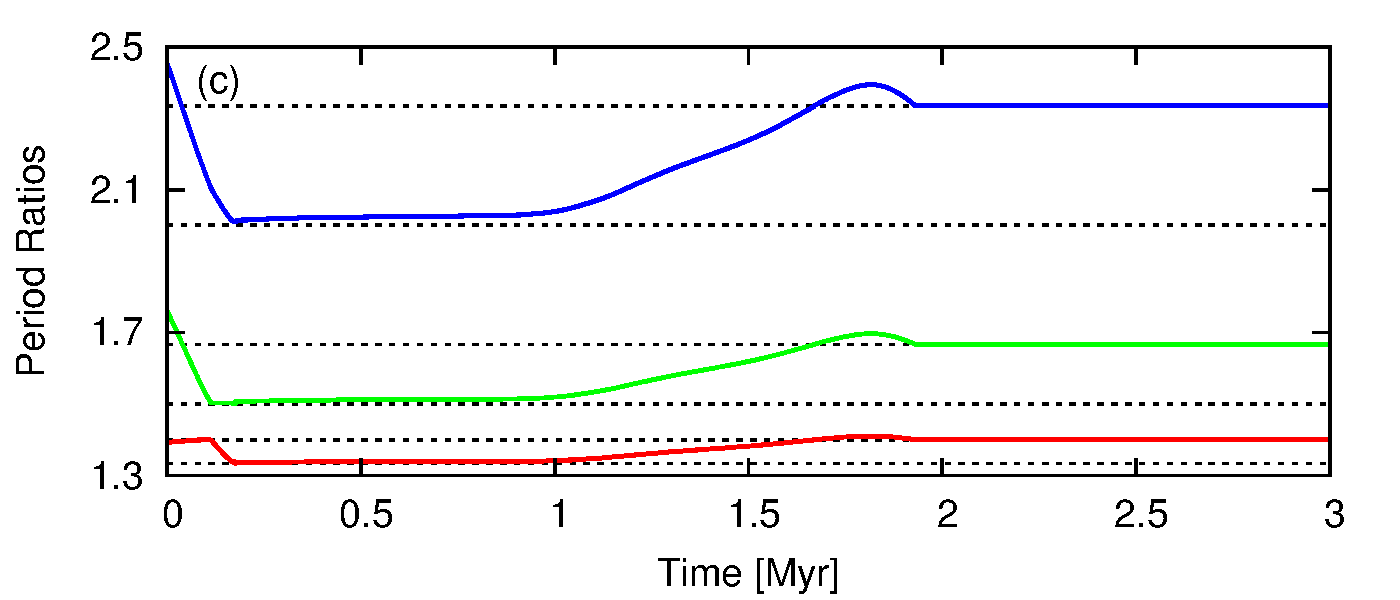}
\includegraphics[width=0.48\textwidth]{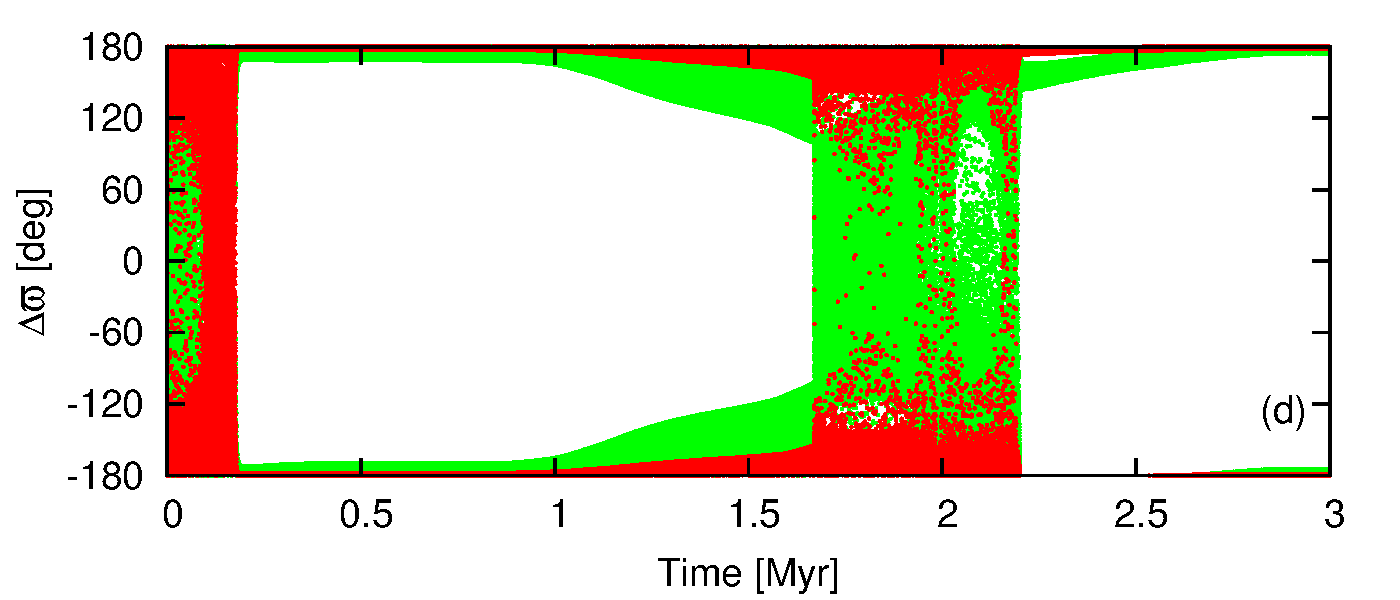}
}
\hbox{
\includegraphics[width=0.48\textwidth]{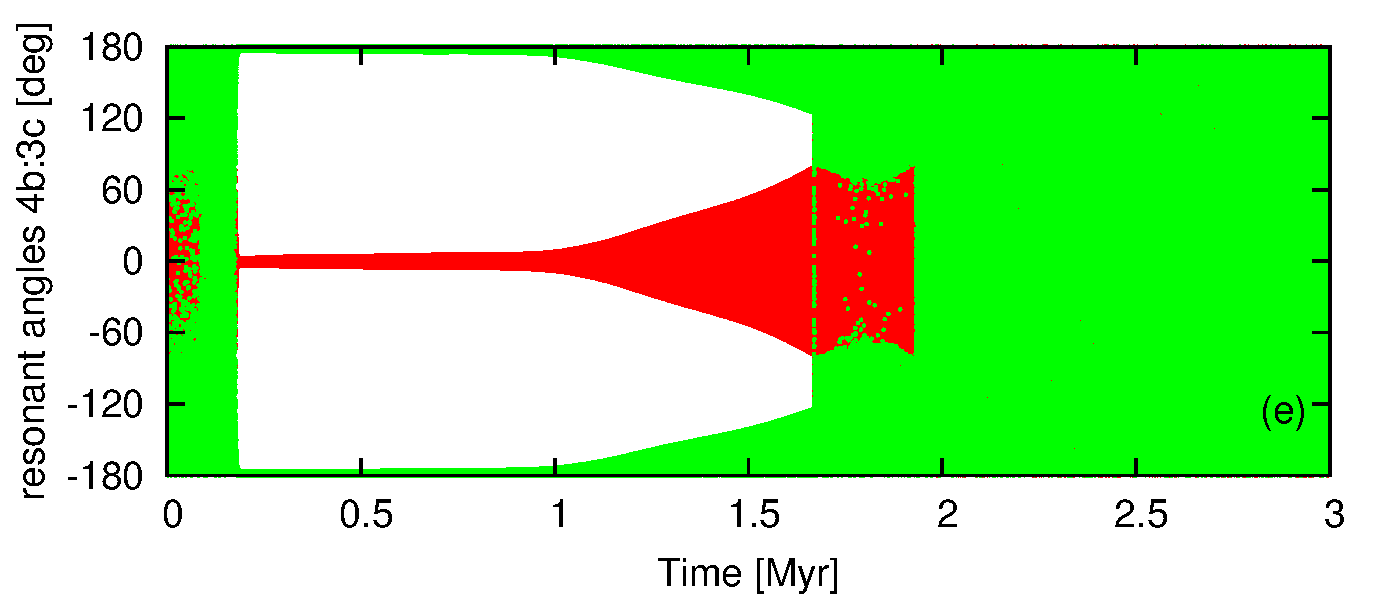}
\includegraphics[width=0.48\textwidth]{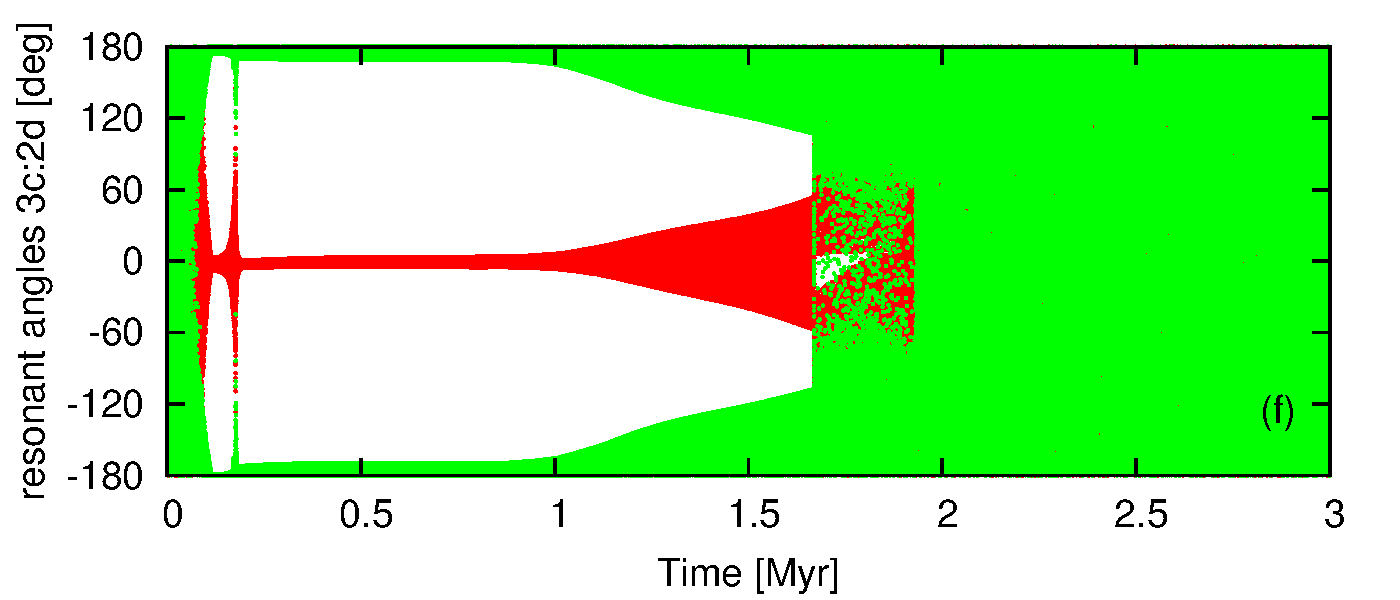}
}
\hbox{
\includegraphics[width=0.48\textwidth]{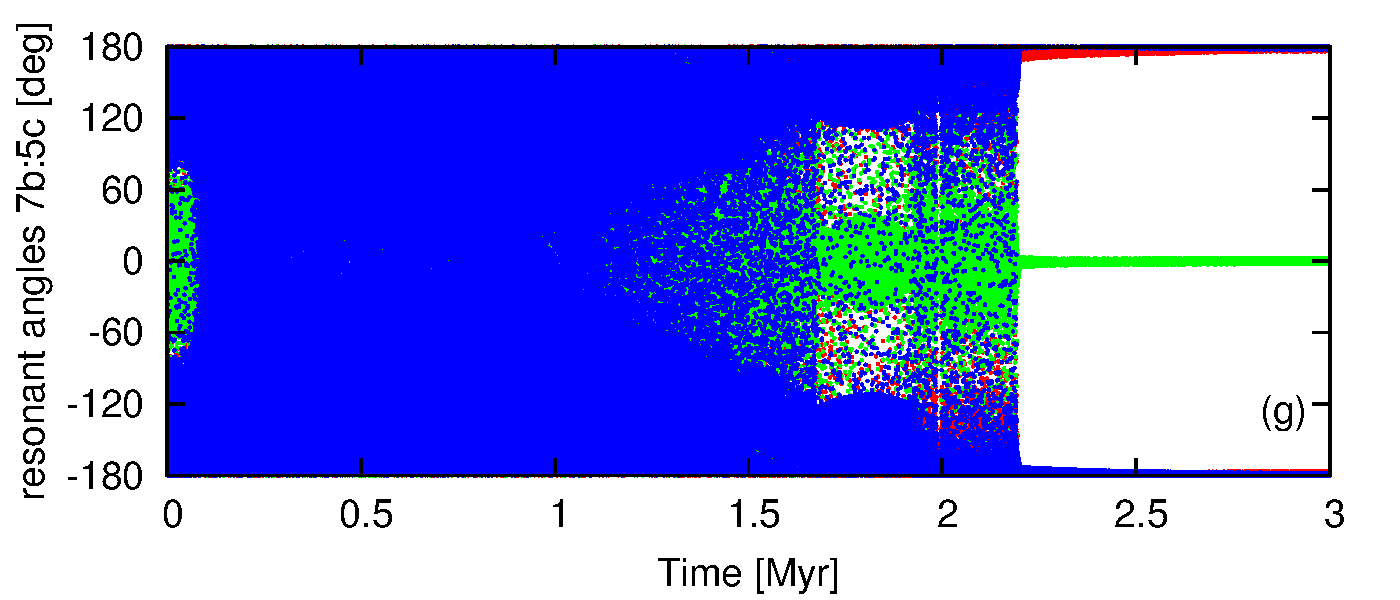}
\includegraphics[width=0.48\textwidth]{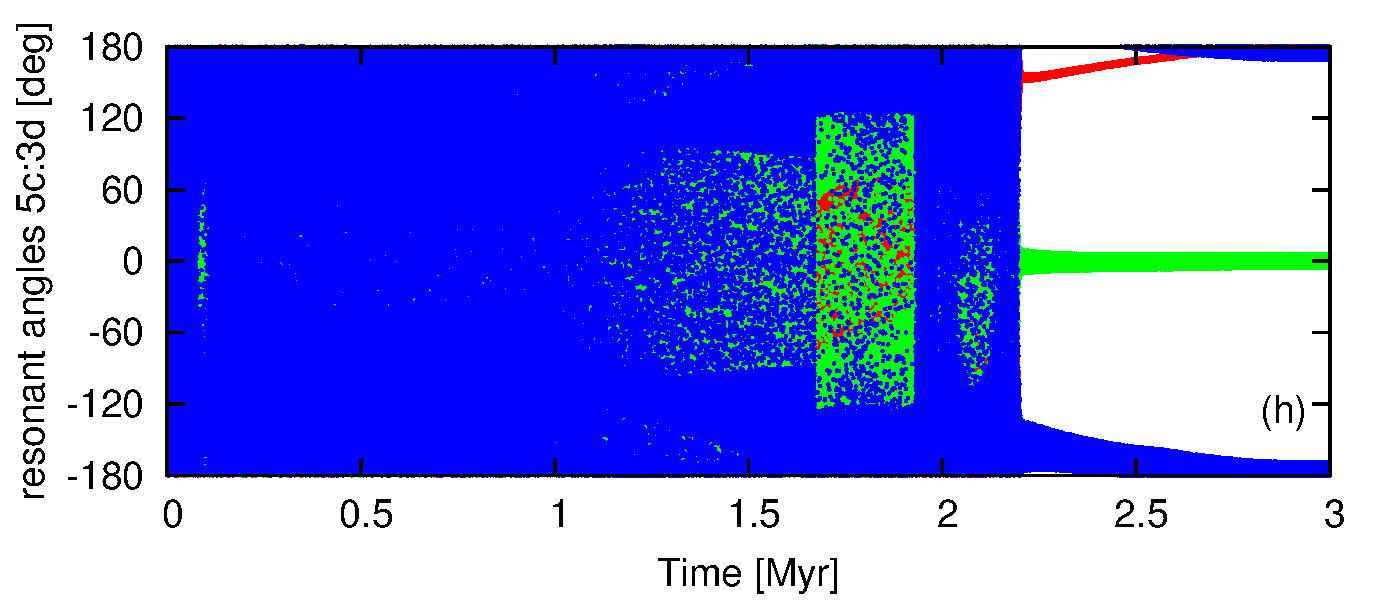}
}
}
\end{center}
\caption{The same as in Fig.~\ref{fig:ex1} but for a system which ends up as a chain of 7:5 and 5:3~MMRs. Planets' masses are $m_1=4.590\,\mE$, $m_2=4.664\,\mE$ and $m_3=4.786\,\mE$.}
\label{fig:ex2}
\end{figure*}

\subsection{Chains of higher-order resonances}

Figure~\ref{fig:ex2} presents the evolution of a system which ends up as a chain of second-order MMRs, namely 7:5 and 5:3 for the inner and the outer pair, respectively. The evolution is shown in the same manner as in Fig.~\ref{fig:ex1}, however for second-order resonances there are three critical angles instead of two, as for the first-order MMRs. The critical angles for 7:5~MMR (for the inner pair) read $\phi_{1,1} = 5\,\lambda_1 - 7\,\lambda_2 + 2\,\varpi_1$, $\phi_{1,2} = 5\,\lambda_1 - 7\,\lambda_2 + \varpi_1 + \varpi_2$, $\phi_{1,3} = 5\,\lambda_1 - 7\,\lambda_2 + 2\,\varpi_2$. The second pair of planets ends up in 5:3~MMRs, thus the angles read $\phi_{2,1} = 3\,\lambda_2 - 5\,\lambda_3 + 2\,\varpi_2$, $\phi_{2,2} = 3\,\lambda_2 - 5\,\lambda_3 + \varpi_2 + \varpi_3$, $\phi_{2,3} = 3\,\lambda_2 - 5\,\lambda_3 + 2\,\varpi_3$.

\begin{figure}
\begin{center}
\vbox{
\hbox{
\includegraphics[width=0.48\textwidth]{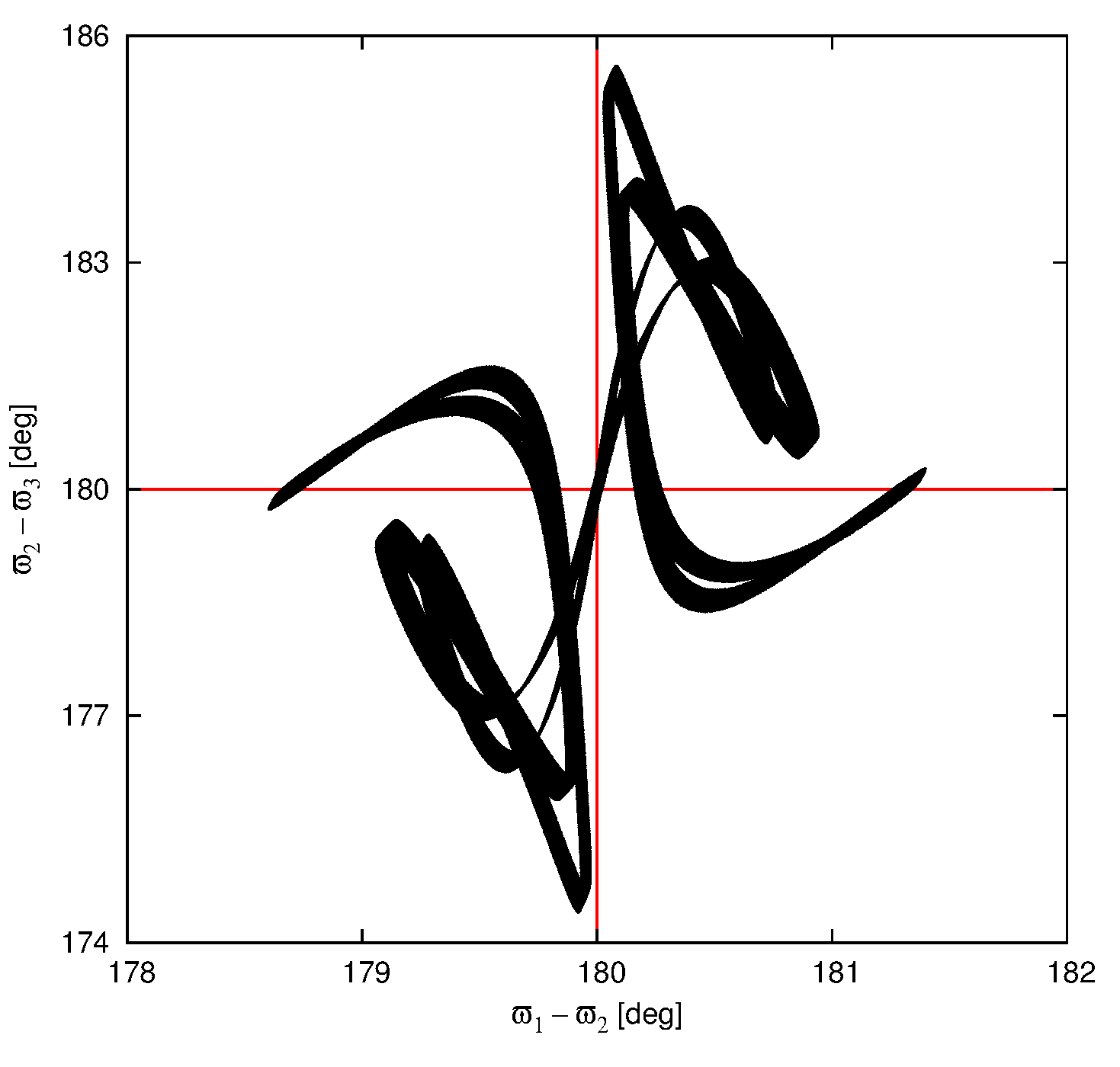}
}
}
\end{center}
\caption{The evolution of the system illustrated in Fig.~\ref{fig:ex2}, after the disc disperses, presented at the plane of $(\Delta\varpi_{1,2}, \Delta\varpi_{2,3})$. The evolution of the system has been integrated over $0.1$~Myr.}
\label{fig:second_order_periodic}
\end{figure}

The system is first locked in a chain of 4:3 and 3:2~MMRs (see panels e and~f of Fig.~\ref{fig:ex2}). At $t \sim 1~$Myr the migration starts to be divergent. The period ratios increase above $7/5$ and $5/3$. Shortly after that the migration is convergent again and the system is locked in a chain of 7:5 and 5:3~MMRs. Final configuration has all critical angles librating, nevertheless the evolution is not exactly periodic but close to periodic (see Fig.~\ref{fig:second_order_periodic}). As the formation of the chain occurred at relatively late stage of the migration, the migration was already slow and the exact periodic configuration was not achieved.

\section{A comparison with the observations}

Figure~\ref{fig:stat2} shows a comparison between synthetic systems (black dots) and observed three-planet systems\footnote{Parameters of the observed systems were taken from the NASA Exoplanet Archive, http://exoplanetarchive.ipac.caltech.edu, Q1-Q17 results were used. There are $82$ systems with three confirmed planets in the sample.} (green dots) at the period ratio -- period ratio diagram. Grey squares show ranges from which the initial orbits for the simulations were chosen. The correspondence between the observed and simulated systems is not clear. The most visible discrepancies are for $P_2/P_1 \in [1.8, 2.2]$ and $P_3/P_2 \in [1.6, 2]$ as there is a gap within this area of the diagram for the synthetic systems sample, while there are 7 systems observed within this range. A good agreement between theory and observation can be observed in the range where both $P_2/P_1 > 2$ and $P_3/P_2 > 2$. The range of both $P_2/P_1 < 2$ and $P_3/P_2 < 2$ needs closer inspection in order to make conclusions.

Naturally, the final distribution of the systems resulting from the simulations depends on the initial distribution as well as the planets' masses and the disc properties. Therefore, we do not present here histograms of the period ratios, which are commonly used to compare the distributions of the period ratios of the synthetic and observed systems. Instead, in the next section we show that the comparison of the observed and the simulated systems at the period ratio -- period ratio plane enables us to put some constrains on the planet-disc interactions model.

\subsection{Divergent evolution of a system in a chain of MMRs}

One of the conclusions of \citep{Batygin2013} is that most of the observed systems with period ratios relatively close but not exactly equal to the resonant values may be in fact resonant (in terms of the resonant angles librations). Such systems could have been closer to exact commensurability in the past but due to the orbital circularization (which is a result of the tidal interaction between the planets and their parent stars), they evolved divergently and moved away from the nominal values of MMRs. This is also a conclusion of Paper~I, although we showed that the divergent migration is relatively common in protoplanetary discs, when a system of two planets of similar masses is considered. The conclusion is reasonable when one looks at the evolution of the system at the period ratio axis, as a system whose period ratios are different from the exact commensurability could have been shifted away from the MMR in either way mentioned above. The situation is different, though, when a three-planet system is considered and its evolution is being studied at the period ratio -- period ratio diagram.

As it was mentioned earlier in this work, \cite{Papaloizou2015} considered the orbital circularization of a three-planet system initially involved in a chain of first-order MMRs, $P_2/P_1 \approx (q+1)/q$ and $P_3/P_2 \approx (p+1)/p$. If the orbits undergo the circularization due to the tidal star-planet interaction, the period ratios $P_2/P_1$ and $P_3/P_2$ increase and so do the amplitudes of the resonant angles librations, although the angles keep librating even for period ratios significantly different from the nominal values of MMRs. The mean motions vary in a way that the Laplace condition is fulfilled, i.e., $(q+1)\,n_2 - q\,n_1 = (p+1)\,n_3 - p\,n_2$. The above implies that the system evolves at the period ratio -- period ratio diagram along a curve $y(x)$ of a form:
\begin{equation}
\frac{1}{y} = 1 + C\,(1 - x), \quad C = \frac{q}{p+1}, \quad x \equiv \frac{P_2}{P_1}, \quad y \equiv \frac{P_3}{P_2}.
\label{eq:yx}
\end{equation}

\begin{figure*}
\begin{center}
\vbox{
\includegraphics[width=0.48\textwidth]{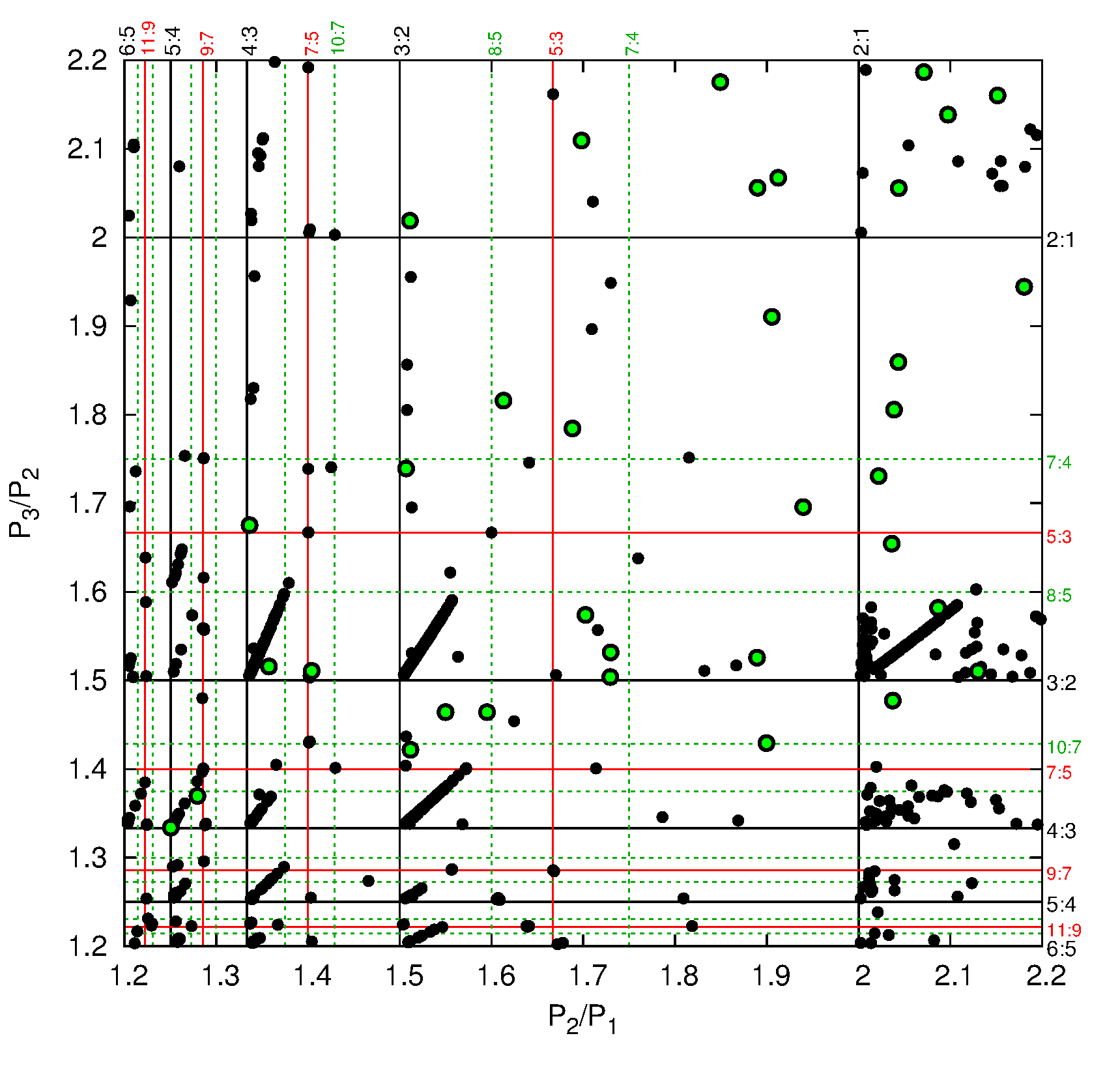}
\includegraphics[width=0.48\textwidth]{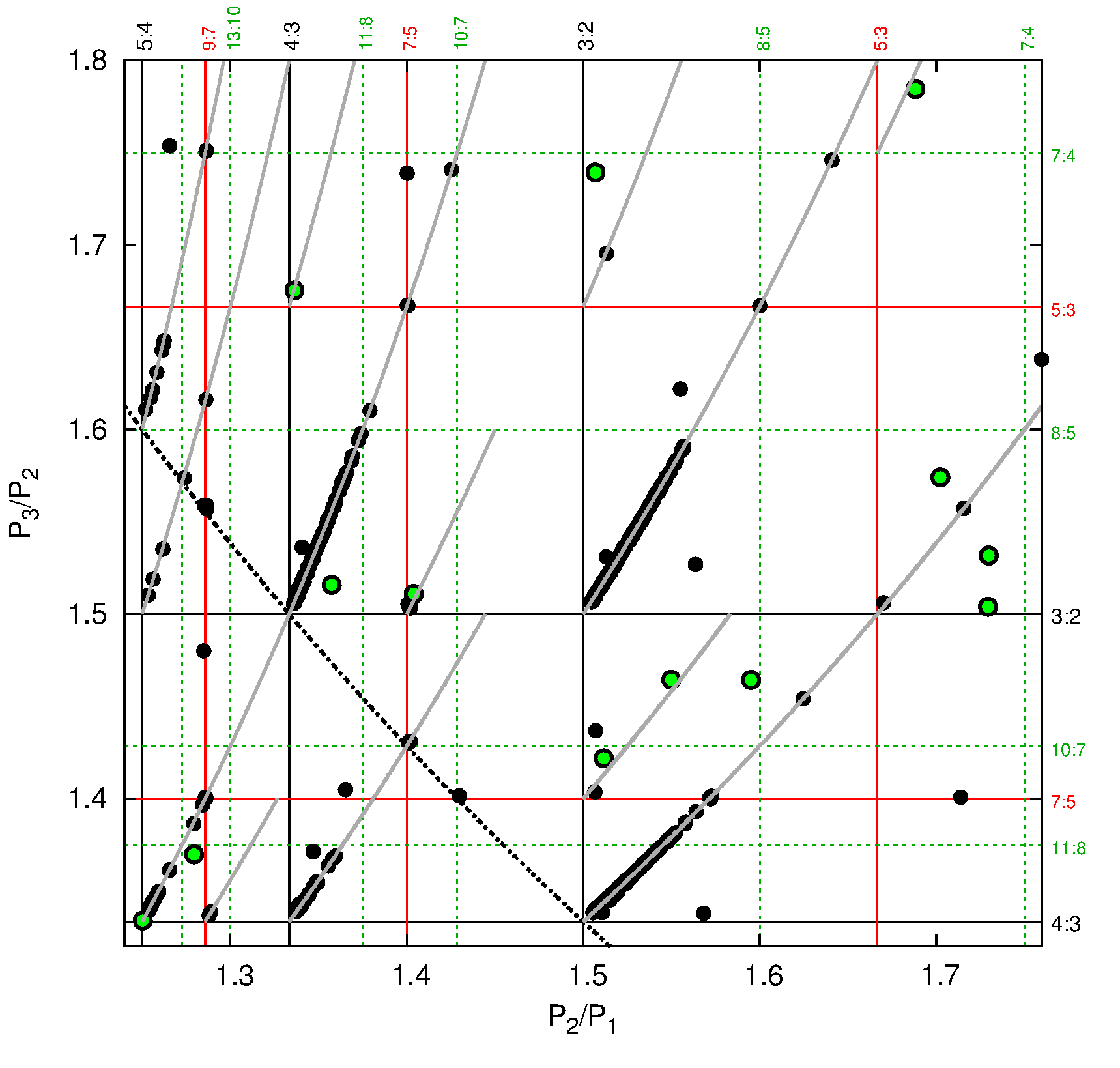}
}
\end{center}
\caption{{\em Left-hand panel:} The synthetic (black dots) and the observed systems (green dots; grey dots in the printed version) presented at $(P_2/P_1, P_3/P_2)-$plane. Vertical and horizontal lines show positions of MMRs (labelled accordingly). {\em Right-hand panel:} A Zoom-in of the left-hand panel. Grey curves show the resonant divergent migration paths for different chains of MMRs.}
\label{fig:stat3}
\end{figure*}

Therefore, if the scenario of resonant systems which evolved divergently from MMRs but kept their critical angles librating was true, the observed systems should lie on the curves defined above (there is one curve for a given chain of MMRs, parametrized by $q$ and $p$ values). Nevertheless, Figure~\ref{fig:stat3} shows that this is not the case. Grey curves at the right-hand panel of Fig.~\ref{fig:stat3} show the resonant divergent migration paths (or tracks). Green dots, showing the positions of the observed systems seem to avoid those paths. On contrary, many black symbols (which mark the positions of the simulated systems) lie on the paths. Similarly to the tracks of the systems deviating from chains of the first-order MMRs, one can find formulae $y(x)$ for higher order MMRs. The general form of $y(x)$ is given by Eq.~\ref{eq:yx}, although the  constant coefficients are different than for the chains of the first-order MMRs. For a chain of $i$-th and $j$-th order MMRs, i.e., $P_2/P_1 \approx (q+i)/q$ and $P_3/P_2 \approx (p+j)/p$, the coefficient reads $C = j\,q/(i\,(p+j))$.

\subsection{Constrains on the $\kappa$ parameter}

The fact that the observed systems whose period ratios are far from exact commensurability do not follow the resonant divergent migration tracks can be explained with several scenarios. 1) Their divergent evolution was perturbed so the systems deviated away from the tracks. 2) The systems never resided in chains of MMRs in a sense that the period ratios were not both close to the nominal values of MMRs. 3) There are more planets in the systems classified as three-planet configurations, which makes the evolution of a given system projected into the $(P_2/P_1, P_3/P_2)-$plane more complex than for a three-planet case. 

\begin{figure*}
\begin{center}
\includegraphics[width=0.7\textwidth]{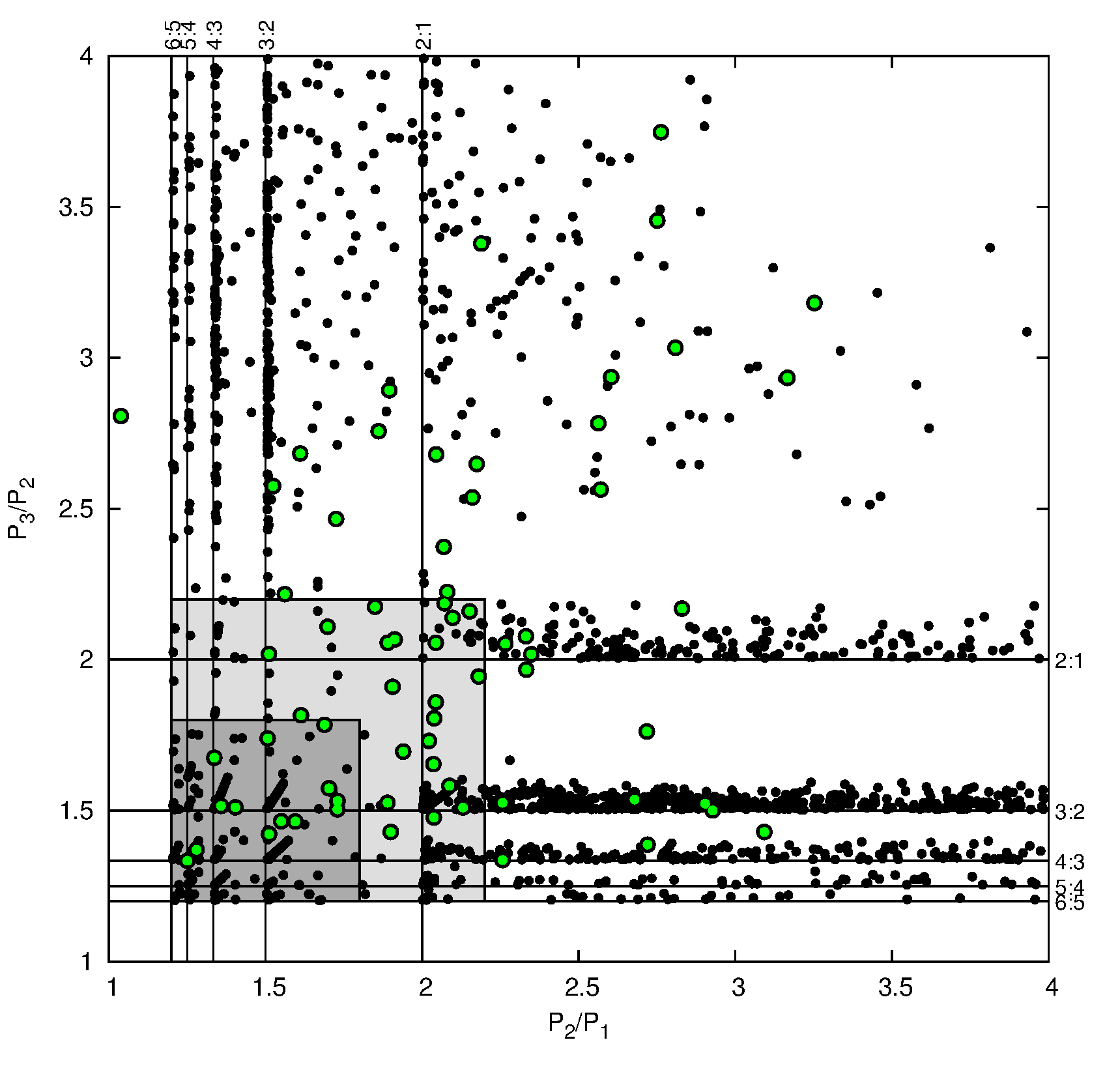}
\end{center}
\caption{The synthetic (black dots) and the observed systems (green dots; grey dots in the printed version) presented at $(P_2/P_1, P_3/P_2)-$plane. Vertical and horizontal lines show positions of first-order MMRs (labelled accordingly). Grey squares show the ranges from which the initial orbits for the simulations were chosen (see the text for details).}
\label{fig:stat2}
\end{figure*}

It is not possible to exclude the last possibility without better observational constraints. The second scenario requires that the migration of a given system was divergent for most of the time when the disc existed, because for a convergent migration chains of resonances are being formed easily. If so, period ratios should be in general high, i.e., if a given system was initially close to the left-bottom corner of Fig.~\ref{fig:stat2}, it should evolve towards the right-top corner of this plot and possibly beyond it. On contrary, we observe systems with both period ratios small ($\lesssim 2$). There are $\sim 60\,\%$ of such systems in the observational sample. It is possible, though, that the planets migrated at very similar rates, i.e., the period ratios did not change much during the migration, even though individual periods changed significantly.

\begin{figure}
\begin{center}
\vbox{
\hbox{
\hspace{-1mm}
\includegraphics[width=0.24\textwidth]{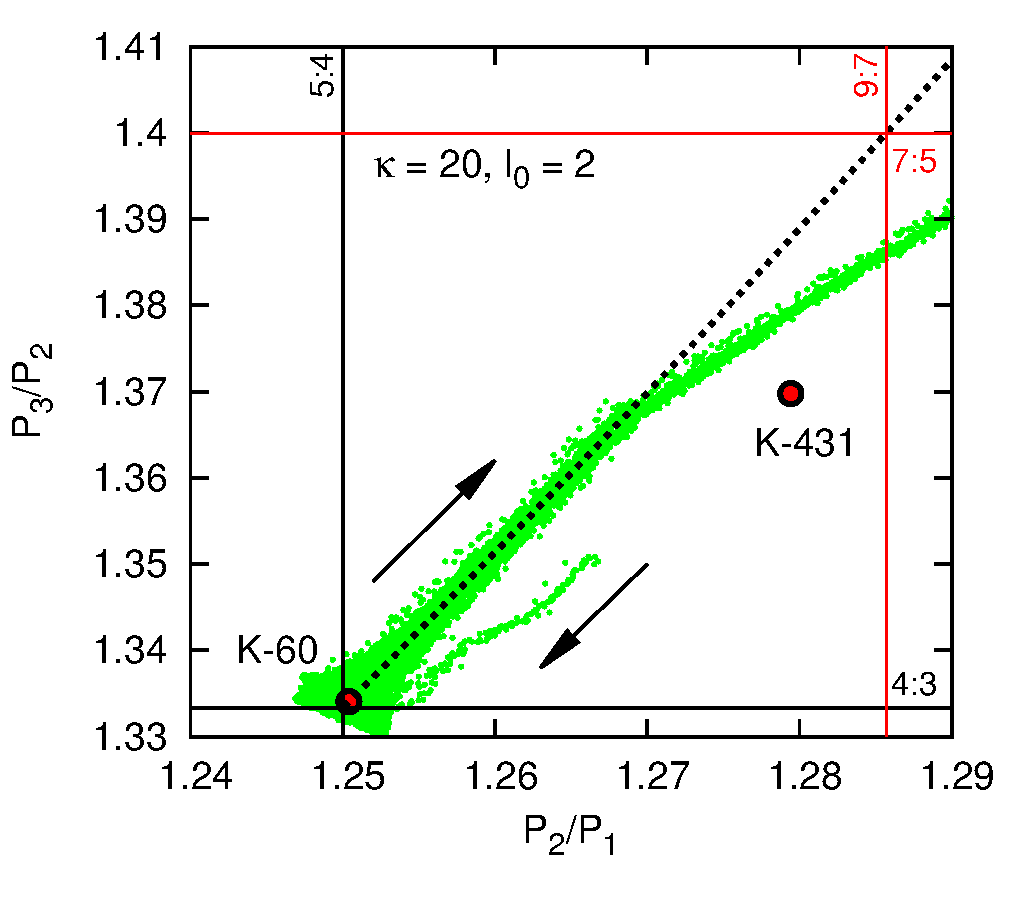}
\hspace{-2mm}
\includegraphics[width=0.24\textwidth]{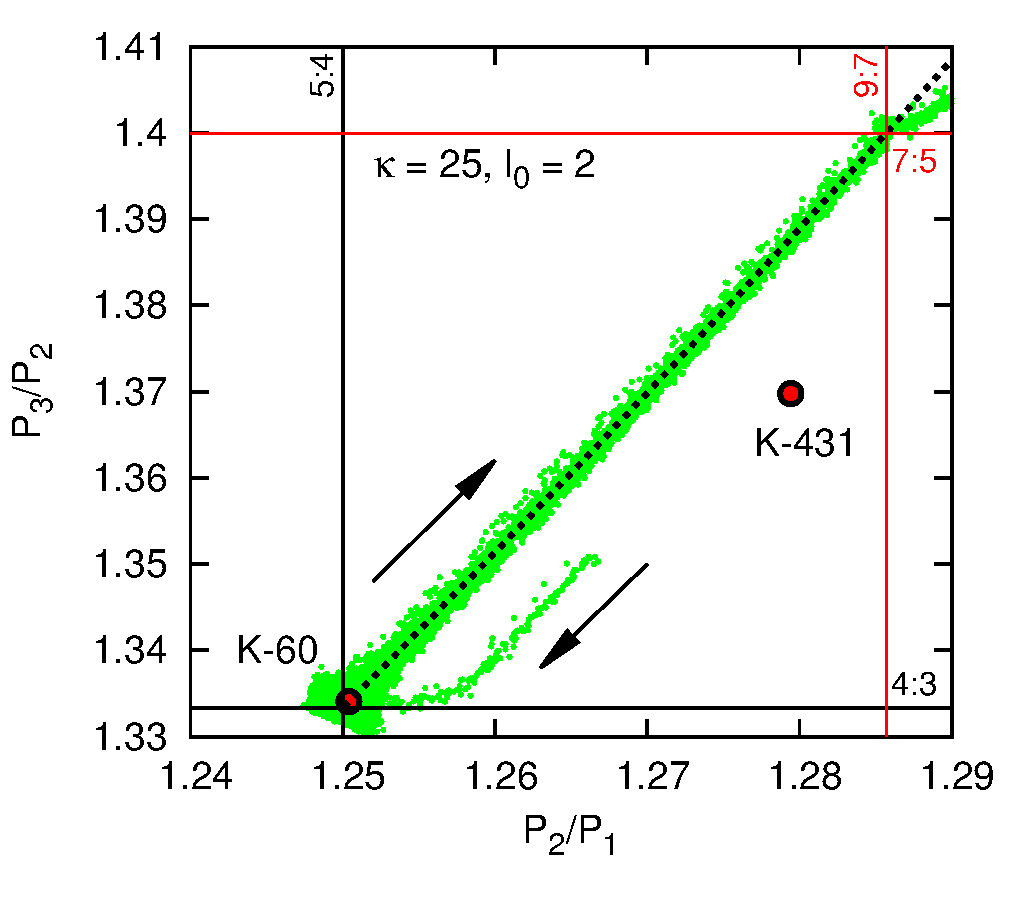}
}
\hbox{
\hspace{-1mm}
\includegraphics[width=0.24\textwidth]{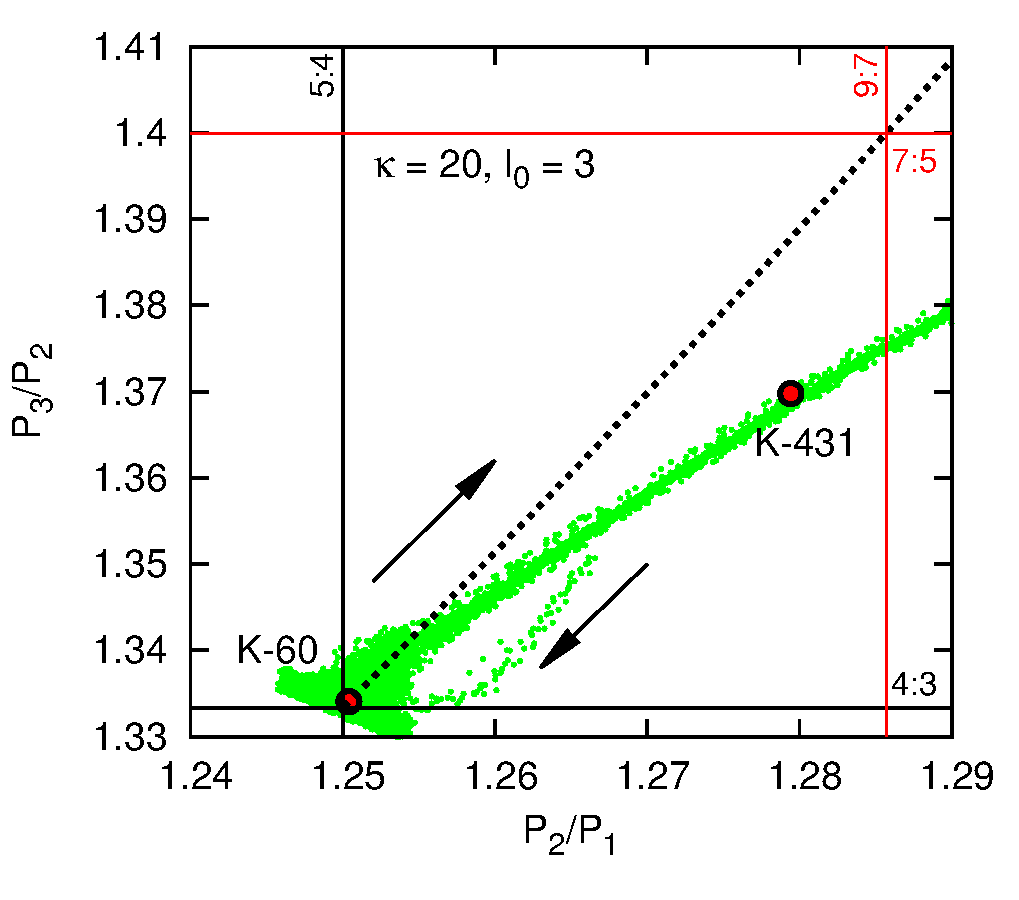}
\hspace{-2mm}
\includegraphics[width=0.24\textwidth]{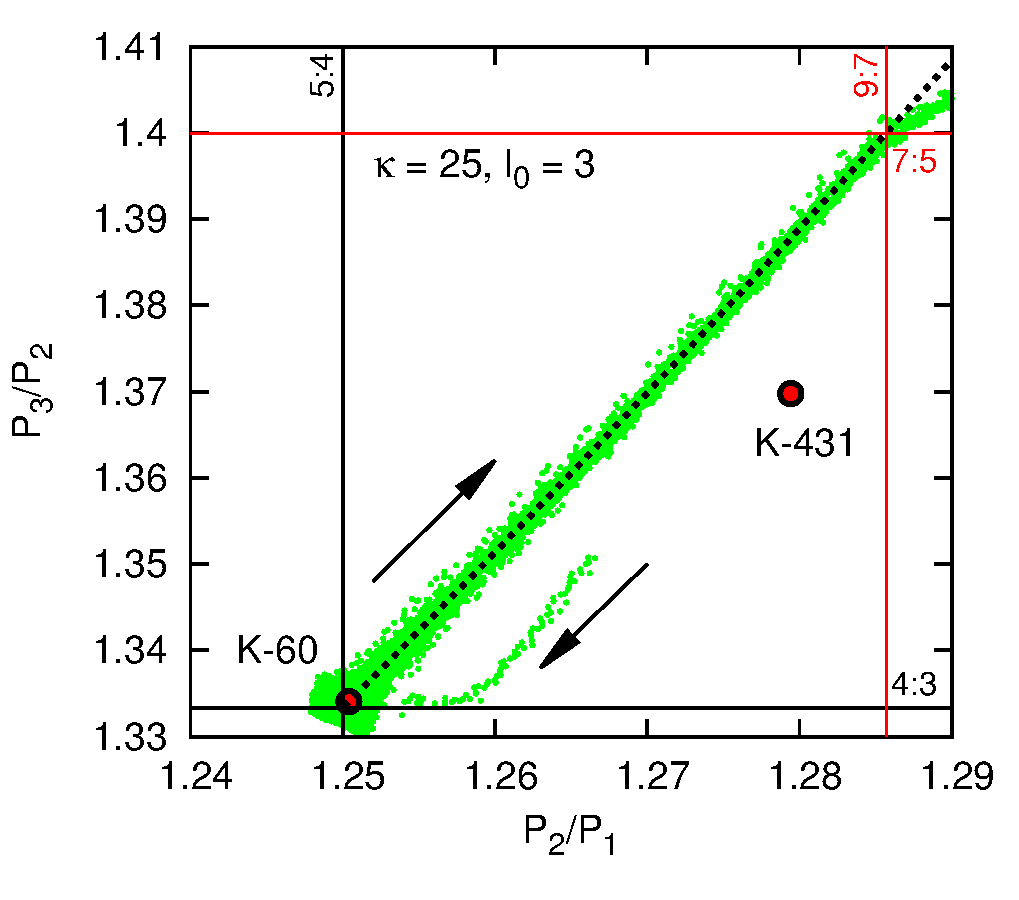}
}
}
\end{center}
\caption{Green dots (light grey in the printed version) at each panel shows the evolution of a chosen system in the realm of a phenomenological model of migration, Eq.~\ref{eq:accel}. Arrows show the direction of the evolution. Vertical and horizontal lines show positions of MMRs (labelled accordingly). The values of the migration parameters (see Eq.~\ref{eq:tau}) are: $\tau_0 = 10$~Myr, $T = 10$~Myr, $l$ varies between $l_0$ and $-l_0$. Each panel presents the evolution for given $\kappa, l_0$ (labelled accordingly). The masses of all three planets are equal, $m_1 = m_2 = m_3 = 3\,\mE$. Red filled circles (grey in the printed version) show positions of Kepler-60 and Kepler-431 systems.}
\label{fig:stat4}
\end{figure}

Lets consider now the first scenario. The system initially migrates convergently and moves towards a chain of MMRs. When both period ratios reach the nominal values of MMRs, the system is locked in the chain. After some time, the migration becomes divergent. If the reason for the divergent migration is the tidally induced circularization, the system should move along curves $y(x)$ defined above. The system remains resonant until it reaches another resonance. Therefore, systems located at the period ratio -- period ratio diagram within a rectangle defined by given low-order MMRs (see Fig.~\ref{fig:stat4} and a rectangle defined by resonances 5:4 and 9:7 for the first pair of planets and 4:3 and 7:5 for the second pair) should be resonant and should remain at the resonant divergent migration path.

In order to illustrate the divergent evolution of a system and to discuss how it could leave the resonant divergent migration path, we consider a simple parametric model of the migration \citep[e.g.,][]{Beauge2006,Moore2013} in which the acceleration $\vec{f}_i$ of an $i$-th planet resulting from the planet-disc interactions reads:
\begin{equation}
\vec{f}_i = -\frac{\vec{v}_i}{2\,\tau_{\idm{a},i}} - \frac{\vec{v}_i - \vec{v}_{\idm{c},i}}{\tau_{\idm{e},i}},
\label{eq:accel}
\end{equation}
where $\vec{v}_i$ is the astrocentric velocity of a given $i$-th planet, $\vec{v}_{\idm{c},i}$ is the Keplerian velocity of that planet in a circular orbit of radius $r_i = \|\vec{r}_i\|$. The time-scales of migration and circularization are given by $\tau_{\idm{a},i}$ and $\tau_{\idm{e},i}$, respectively. In general they can be arbitrary functions of planets' masses, positions, velocities and time. In order to show correspondence between the above model and the one used by \cite{Papaloizou2015} and \cite{Batygin2013} we write down formulae for the semi-major axes and the eccentricities evolution of a single-planet averaged out over the orbital motion, when $\tau_a$ and $\tau_e$ are constant. We have:
\begin{eqnarray}
\dot{a} &=& -\frac{a}{\tau_a} \left( 1 + \frac{5}{8} \, e^2 \, \kappa + O(e^4)\right), \\
\dot{e} &=& -\frac{e}{\tau_e} \left( 1 - \frac{13}{32}\, e^2 + O(e^4) \right),
\end{eqnarray}
where $\kappa \equiv \tau_a/\tau_e$. 
A model studied in the cited papers can be characterized by $\kappa \to \infty$ and finite $\tau_e$, thus $\dot{a} = -(5/8) \, a \, e^2 \, \tau_e + O(e^4)$, although they have a different constant factor ($2$ instead of $5/8$), thus the interpretation of $\tau_e$ is slightly different.

In order to follow the scenario outlined above (first convergent, then divergent migration) we chose the following functional form of $\tau_a$ (we chose $\kappa$ to be constant):
\begin{equation}
\tau_{a,i} = \tau_0 \left( \frac{r_i}{1\,\au} \right)^{-l(t)} \, \exp(t/T),
\label{eq:tau}
\end{equation}
where $\tau_0 = 10~$Myr, $T = 10~$Myr and $i=1,2,3$. The migration is convergent if $l>0$ and divergent if $l<0$. At the beginning of the simulation $l=l_0$, it decreases linearly in time such that $l=0$ at $t=5~$Myr (all planets migrate at the same rate), after another $5~$Myr the power index decreases down to $l=-l_0$ and remains constant later on. We performed simulations for different values of $l_0$ and $\kappa$, the results of which are presented in Fig.~\ref{fig:stat4}. The system starts from close-to-circular orbits of sizes of $a_1 = 1\,\au$, $a_2=1.17\,\au$ and $a_3=1.43\,\au$, which give the period ratio slightly above the resonant values of 5:4 and 4:3 for the inner and outer pair, respectively. A given system can leave the resonant divergent migration path when a value of $\kappa$ is relatively low, i.e., of the order of $20$ or below. In such an instance, the system is not resonant any more. An example of a system which could be formed this way is Kepler-431 (whose position is shown with a red symbol in Fig.~\ref{fig:stat4}).
On contrary, Kepler-60 system, which was recently shown to be involved in a chain of resonances \citep{Gozdziewski2016} unlikely experienced divergent migration after it was trapped in a chain of 5:4, 4:3~MMRs.

Using formulae for $\tau_a$ and $\tau_e$ from \citep{Tanaka2002,Tanaka2004}, which were obtained for the isothermal disc model with temperature profiles given by power law in radii with the index of $1/2$, one obtains $\kappa = 160 \, ( r/1\,\au )^{-1/2}$.
The model used in our work leads to even higher values of $\kappa$. For orbits inside $1\,\au$ $\kappa$ is typically of the order of a few hundreds up to $\sim 1000$. 

Hydrodynamical simulations lead to similar values of the circularization time-scale as provided by the formula from \citep{Tanaka2004} for eccentricities smaller than the aspect ratio of the disc, $h$, \citep{Cresswell2007,Bitsch2010}. For the radiative disc model \cite{Bitsch2010} found that the circularization may be slower, but only by a factor of $3$, when compared to the results obtained for the isothermal disc. However, in other papers \citep[e.g.,][]{Papaloizou2000,Cresswell2006} one can find the hydrodynamical simulations which lead to even higher circularization rates (by an order of magnitude) than the results of \cite{Tanaka2004}. Particular  estimations of $\kappa$ were compared in \citep{Muto2011,Ketchum2011}, nevertheless a way to decrease $\kappa$ is still unknown. On the other hand, \cite{Kley2004} showed that hydrodynamical model of the disc can lead to $\kappa$ as small as $1$, although in this paper a system of Jovian planets' in orbits of moderate ($\gtrsim 0.1$) eccentricities was considered and, as the authors conclude, the disc model might have been too simple. In a newer paper \citep{Kley2009} the simulations in 3D viscous radiative disc lead to $\kappa \sim 30$ for a planet of $20\,\mE$ at moderately eccentric orbit. Nevertheless, the deviation from the path happen when the eccentricities are small, thus the fact that $\kappa(e)$ is lower for higher $e$ \citep[see][]{Papaloizou2000,Cresswell2008} would not change the outcome of the divergent migration.

On the other hand, low-$\kappa$ divergent migration is not the only possibility to form systems like Kepler-431. Turbulences in the disc \citep{Nelson2005} as well as interactions with planetesimals \citep{Chatterjee2014} could lead to a similar effect. On the other hand, such stochastic forces acting on the planets could probably make it difficult to form resonant systems with period ratios very close the nominal values of MMRs, like the Kepler-60 system.

\section{Conclusions}

We have studied the migration of three super-Earth mass planets embedded in a protoplanetary disc (see Paper~I for the disc model details). The planets' masses and initial orbits were chosen randomly from ranges defined in Section~2. About one third of the whole sample of $2700$ systems ended up as compact configurations, which we roughly define as systems with both period ratios smaller than $\sim 2.2$. We found that most of the compact systems are involved in chains of MMRs in terms of the critical angles librations. Nevertheless, the angles librate not only for systems with period ratios close to the nominal values of the resonances, but also for systems far from exact commensurability. 

A given system may leave a chain of MMRs in a sense of the period ratios if the migration is divergent. We showed that for $\kappa \gtrsim 25$ systems which undergo the divergent migration remain resonant. Resonant systems which migrate divergently have to follow certain paths at the diagram of the period ratios (which we call the resonant divergent migration paths). For lower values of $\kappa$ a system can leave the path and migrate along a track defined by individual values of $\tau_{1,2}$ and $\tau_{2,3}$. In such an instance the system is not resonant. 

Because most of the observed systems, whose period ratios are shifted with respect to the nominal values of MMRs, do not lie on the resonant divergent migration paths and $\kappa \gtrsim 25$ for typical discs, we conclude that those systems were probably not formed as a result of divergent migration out of the nominal chain of MMRs a given system resided in before. 

We showed that the evolution of three planets migrating in a disc occurs (under certain conditions discussed in the text) along families of periodic orbits. As a result, similarly to a two-planet case, a typical outcome of the migration of three planets, whose period ratios are small, is a periodic configuration. It occurs for chains of first-order MMRs as well as chains of higher order resonance.

The discrepancy between the results of simulations and the observed systems indicates that processes other than smooth migration (turbulences, interactions with planetesimals) likely play a non-negligible role, at least for planets in the Earth to super-Earth mass regime. Another key-factor in the studies of migration is the circularization rate, which is still an open problem. As it was shown in this work, periodic orbits play a crucial role in the migration induced formation of mean motion resonances. Unlike for the two-planet systems, the periodic orbits of chains of resonances have not been studied in the literature. We hope to address all the problems mentioned above in our future works.

\section*{acknowledgements}

I would like to thank the anonymous reviewer for comments which helped to improve the manuscript. 
This work was supported by Polish National Science Centre MAESTRO grant DEC-2012/06/A/ST9/00276. The computations were performed on HPC cluster HAL9000 of the Computing Centre of the Faculty of Mathematics and Physics at the University of Szczecin.

\bibliographystyle{mn2e}
\bibliography{ms}
\label{lastpage}
\end{document}